\documentclass{article}

    \PassOptionsToPackage{numbers, compress}{natbib}

    \usepackage[preprint]{neurips_2025}

\usepackage[utf8]{inputenc} 
\usepackage[T1]{fontenc}    
\usepackage{hyperref}       
\usepackage{url}            
\usepackage{booktabs}       
\usepackage{amsfonts}       
\usepackage{nicefrac}       
\usepackage{microtype}      
\usepackage{xcolor}         

\usepackage{graphicx}

\usepackage{csquotes}
\usepackage{tabularx}
\usepackage{listings}
\usepackage{adjustbox}
\usepackage[binary-units]{siunitx}
\usepackage{xspace}
\usepackage{microtype}
\usepackage{xspace}
\usepackage{flushend}
\usepackage{paralist}
\usepackage{enumitem}
\usepackage{siunitx}
\usepackage{makecell}
\usepackage{pgf}
\usepackage{soul}
\usepackage{amsmath}
\usepackage{subcaption}
\usepackage{algorithm}
\usepackage{algpseudocode}
\usepackage{xcolor}
\usepackage{hyperref}
\usepackage{pgf}
\usepackage{array}
\usepackage{pifont}
\usepackage{multirow}
\usepackage{footnote}
\usepackage{graphicx}
\usepackage{subcaption}
\usepackage{booktabs}
\usepackage{varwidth}

\usepackage{tikz} 
\usetikzlibrary{shapes.geometric, positioning} 

\usepackage[most]{tcolorbox}   
\usepackage{xcolor}            
\usepackage{lipsum}            
\usepackage{microtype}         

\definecolor{darkblue}{rgb}{0.0, 0.0, 0.55} 
\definecolor{darkgreen}{rgb}{0.0, 0.55, 0.0} 

\newif\ifcomments

\commentstrue

\ifcomments
\providecommand{\AS}[1]{\textbf{\textcolor{blue}{AS: #1}}}
\providecommand{\CD}[1]{\textbf{\textcolor{red}{CD: #1}}}
\else
\providecommand{\CD}[1]{}
\providecommand{\AS}[1]{}
\fi

\newcommand{\eg}{e.g.,~}
\newcommand{\ie}{i.e.~}

\newcommand{\etal}{et al.~}

\newcommand{\crosshair}{\texttt{Crosshair}\xspace}

\newcommand{\gptturbo}{\texttt{gpt-3.5-turbo-instruct}\xspace}
\newcommand{\gptembed}{\texttt{text-embedding-ada-002}\xspace}
\newcommand{\gptcodebabbage}{\texttt{code-search-babbage-001}\xspace}
\newcommand{\gptcodeada}{\texttt{code-search-ada-001}\xspace}
\newcommand{\deepseek}{\texttt{DeepSeek-R1}\xspace}

\newcommand{\gpt}{\textsc{GPT}\xspace}
\newcommand{\claude}{\texttt{claude-3-7-sonnet}\xspace}
\newcommand{\openai}{\textsc{OpenAI}\xspace}

\newcommand{\livecodebench}{\textsc{LiveCodeBench}\xspace}
\newcommand{\codebleu}{\textsc{CodeBLEU}\xspace}

\newcommand{\deberta}{\textsc{DeBERTa}\xspace}

\newcommand{\SE}{\textsc{SE}\xspace}
\newcommand{\MI}{\textsc{MI}\xspace}
\newcommand{\CC}{\textsc{CC}\xspace}

\newcommand{\NLG}{\textsc{NLG}\xspace}
\newcommand{\codeembed}{\textsc{CodeEmbed}\xspace}
\newcommand{\symb}{\textsc{Symb}\xspace}
\newcommand{\symbuniform}{\textsc{SymbUniform}\xspace}

\newcommand{\SEOriginal}{\textsc{SE-NLG}\xspace}

\newcommand{\SECodeBLEU}{\textsc{SE-CodeBLEU}\xspace}
\newcommand{\SEEmbed}{\textsc{SE-CodeEmbed}\xspace}

\newcommand{\SESymbolic}{\textsc{SE-SymbNorm}\xspace}
\newcommand{\SESymbolicUnif}{\textsc{SE-SymbUniform}\xspace}

\newcommand{\MIOriginal}{\textsc{MI-NLG}\xspace}
\newcommand{\MICodeBLEU}{\textsc{MI-CodeBLEU}\xspace}
\newcommand{\MIEmbed}{\textsc{MI-CodeEmbed}\xspace}

\newcommand{\MISymbolicUnif}{\textsc{MI-SymbUniform}\xspace}

\newcommand{\MISymbolic}{\textsc{MI-SymbNorm}\xspace}

\newcommand{\LLMProbability}{\textsc{LLM-Probability}\xspace}

\newcommand{\CCSymbolic}{\textsc{CC-Symb}\xspace}
\newcommand{\CCEmbedding}{\textsc{CC-CodeEmbed}\xspace}

\newcommand{\totalProbsComb}{\SI{831}{}\xspace} 
\newcommand{\totalProbs}{\SI{239}{}\xspace} 
\newcommand{\totalProbsMedium}{\SI{332}{}\xspace} 
\newcommand{\totalProbsHard}{\SI{260}{}\xspace} 

\newcommand{\CorrectnessTimeout}{\SI{5}{s}\xspace} 
\newcommand{\CrosshairPerConditiontimeout}{\SI{5}{s}\xspace} 
\newcommand{\CrosshairPerPathtimeout}{\SI{5}{s}\xspace} 
\newcommand{\Crosshairtotaltimeout}{\SI{10}{s}\xspace}

\newcommand{\numSamples}{\SI{5}{}\xspace}
\newcommand{\numIterations}{\SI{2}{}\xspace}

\newcommand{\SENLGPearsonGPTUpdated}{\SI{-0.04}{}}
\newcommand{\SENLGPearsonGPTMediumUpdated}{\SI{-0.02}{}}
\newcommand{\SENLGPearsonGPTHardUpdated}{\SI{-0.01}{}}

\newcommand{\SENLGPearsonPValueGPTUpdated}{\SI{0.5792}{}}
\newcommand{\SENLGPearsonPValueGPTMediumUpdated}{\SI{0.6520}{}}
\newcommand{\SENLGPearsonPValueGPTHardUpdated}{\SI{0.2941}{}}

\newcommand{\SENLGPearsonDeepSeek}{\SI{-0.06}{}}
\newcommand{\SENLGPearsonDeepSeekMedium}{\SI{-0.06}{}}
\newcommand{\SENLGPearsonDeepSeekHard}{\SI{-0.08}{}}

\newcommand{\SENLGPearsonPValueDeepSeek}{\SI{0.2762}{}}
\newcommand{\SENLGPearsonPValueDeepSeekMedium}{\SI{0.1651}{}}
\newcommand{\SENLGPearsonPValueDeepSeekHard}{\SI{0.9289}{}}

\newcommand{\SESymbolicPearsonGPTUpdated}{\SI{-0.56}{}}
\newcommand{\SESymbolicPearsonGPTMediumUpdated}{\SI{-0.52}{}}
\newcommand{\SESymbolicPearsonGPTHardUpdated}{\SI{-0.49}{}}

\newcommand{\SESymbolicPearsonPValueGPTUpdated}{\SI{0.0001}{}}
\newcommand{\SESymbolicPearsonPValueGPTMediumUpdated}{\SI{0.0002}{}}
\newcommand{\SESymbolicPearsonPValueGPTHardUpdated}{\SI{0.0002}{}}

\newcommand{\SESymbolicPearsonDeepSeek}{\SI{-0.45}{}}
\newcommand{\SESymbolicPearsonDeepSeekMedium}{\SI{-0.40}{}}
\newcommand{\SESymbolicPearsonDeepSeekHard}{\SI{-0.41}{}}

\newcommand{\SESymbolicPearsonPValueDeepSeek}{\SI{0.0002}{}}
\newcommand{\SESymbolicPearsonPValueDeepSeekMedium}{\SI{0.0001}{}}
\newcommand{\SESymbolicPearsonPValueDeepSeekHard}{\SI{0.0008}{}}

\newcommand{\SEEmbedPearsonGPT}{\SI{-0.10}{}}
\newcommand{\SEEmbedPearsonGPTMedium}{\SI{0.07}{}}
\newcommand{\SEEmbedPearsonGPTHard}{\SI{-0.02}{}}

\newcommand{\SEEmbedPearsonPValueGPT}{\SI{0.1083}{}}
\newcommand{\SEEmbedPearsonPValueGPTMedium}{\SI{0.1867}{}}
\newcommand{\SEEmbedPearsonPValueGPTHard}{\SI{0.1945}{}}

\newcommand{\SEEmbedPearsonDeepSeek}{\SI{-0.09}{}}
\newcommand{\SEEmbedPearsonDeepSeekMedium}{\SI{-0.08}{}}
\newcommand{\SEEmbedPearsonDeepSeekHard}{\SI{-0.08}{}}

\newcommand{\SEEmbedPearsonPValueDeepSeek}{\SI{0.1631}{}}
\newcommand{\SEEmbedPearsonPValueDeepSeekMedium}{\SI{0.1409}{}}
\newcommand{\SEEmbedPearsonPValueDeepSeekHard}{\SI{0.1567}{}}

\newcommand{\SESymbolicUnifPearsonGPTUpdated}{\SI{-0.51}{}}
\newcommand{\SESymbolicUnifPearsonGPTMediumUpdated}{\SI{-0.48}{}}
\newcommand{\SESymbolicUnifPearsonGPTHardUpdated}{\SI{-0.47}{}}

\newcommand{\SESymbolicUnifPearsonPValueGPTUpdated}{\SI{0.0001}{}}
\newcommand{\SESymbolicUnifPearsonPValueGPTMediumUpdated}{\SI{0.0001}{}}
\newcommand{\SESymbolicUnifPearsonPValueGPTHardUpdated}{\SI{0.0002}{}}

\newcommand{\SESymbolicUnifPearsonDeepSeek}{\SI{-0.45}{}}
\newcommand{\SESymbolicUnifPearsonDeepSeekMedium}{\SI{-0.41}{}}
\newcommand{\SESymbolicUnifPearsonDeepSeekHard}{\SI{-0.40}{}}

\newcommand{\SESymbolicUnifPearsonPValueDeepSeek}{\SI{0.0002}{}}
\newcommand{\SESymbolicUnifPearsonPValueDeepSeekMedium}{\SI{0.0001}{}}
\newcommand{\SESymbolicUnifPearsonPValueDeepSeekHard}{\SI{0.0007}{}}

\newcommand{\SESymbolicUnifPearsonClaude}{\SI{-0.35}{}}
\newcommand{\SESymbolicUnifPearsonClaudeMedium}{\SI{-0.34}{}}
\newcommand{\SESymbolicUnifPearsonClaudeHard}{\SI{-0.32}{}}

\newcommand{\SESymbolicUnifPearsonPValueClaude}{\SI{0.0002}{}}
\newcommand{\SESymbolicUnifPearsonPValueClaudeMedium}{\SI{0.0002}{}}
\newcommand{\SESymbolicUnifPearsonPValueClaudeHard}{\SI{0.0003}{}}

\newcommand{\MINLGPearsonGPTUpdated}{\SI{-0.02}{}}
\newcommand{\MINLGPearsonGPTMediumUpdated}{\SI{-0.04}{}}
\newcommand{\MINLGPearsonGPTHardUpdated}{\SI{-0.07}{}}

\newcommand{\MINLGPearsonPValueGPTUpdated}{\SI{0.6353}{}}
\newcommand{\MINLGPearsonPValueGPTMediumUpdated}{\SI{0.1828}{}}
\newcommand{\MINLGPearsonPValueGPTHardUpdated}{\SI{0.4748}{}}

\newcommand{\MINLGPearsonDeepSeek}{\SI{-0.05}{}}
\newcommand{\MINLGPearsonDeepSeekMedium}{\SI{-0.04}{}}
\newcommand{\MINLGPearsonDeepSeekHard}{\SI{-0.05}{}}

\newcommand{\MINLGPearsonPValueDeepSeek}{\SI{0.7674}{}}
\newcommand{\MINLGPearsonPValueDeepSeekMedium}{\SI{0.8372}{}}
\newcommand{\MINLGPearsonPValueDeepSeekHard}{\SI{0.5365}{}}

\newcommand{\MISymbolicPearsonGPTUpdated}{\SI{-0.22}{}}
\newcommand{\MISymbolicPearsonGPTMediumUpdated}{\SI{-0.24}{}}
\newcommand{\MISymbolicPearsonGPTHardUpdated}{\SI{-0.19}{}}

\newcommand{\MISymbolicPearsonPValueGPTUpdated}{\SI{0.0292}{}}
\newcommand{\MISymbolicPearsonPValueGPTMediumUpdated}{\SI{0.0132}{}}
\newcommand{\MISymbolicPearsonPValueGPTHardUpdated}{\SI{0.0145}{}}

\newcommand{\MISymbolicPearsonDeepSeek}{\SI{-0.37}{}}
\newcommand{\MISymbolicPearsonDeepSeekMedium}{\SI{-0.38}{}}
\newcommand{\MISymbolicPearsonDeepSeekHard}{\SI{-0.33}{}}

\newcommand{\MISymbolicPearsonPValueDeepSeek}{\SI{0.0261}{}}
\newcommand{\MISymbolicPearsonPValueDeepSeekMedium}{\SI{0.0121}{}}
\newcommand{\MISymbolicPearsonPValueDeepSeekHard}{\SI{0.0187}{}}

\newcommand{\MISymbolicUnifPearsonGPT}{\SI{-0.21}{}}
\newcommand{\MISymbolicUnifPearsonGPTMedium}{\SI{-0.18}{}}
\newcommand{\MISymbolicUnifPearsonGPTHard}{\SI{-0.19}{}}

\newcommand{\MISymbolicUnifPearsonPValueGPT}{\SI{0.0002}{}}
\newcommand{\MISymbolicUnifPearsonPValueGPTMedium}{\SI{0.0004}{}}
\newcommand{\MISymbolicUnifPearsonPValueGPTHard}{\SI{0.0008}{}}

\newcommand{\MISymbolicUnifPearsonDeepSeek}{\SI{-0.33}{}}
\newcommand{\MISymbolicUnifPearsonDeepSeekMedium}{\SI{-0.34}{}}
\newcommand{\MISymbolicUnifPearsonDeepSeekHard}{\SI{-0.32}{}}

\newcommand{\MISymbolicUnifPearsonPValueDeepSeek}{\SI{0.0001}{}}
\newcommand{\MISymbolicUnifPearsonPValueDeepSeekMedium}{\SI{0.0004}{}}
\newcommand{\MISymbolicUnifPearsonPValueDeepSeekHard}{\SI{0.0003}{}}

\newcommand{\MIEmbedPearsonGPT}{\SI{-0.024}{}}
\newcommand{\MIEmbedPearsonGPTMedium}{\SI{-0.056}{}}
\newcommand{\MIEmbedPearsonGPTHard}{\SI{-0.081}{}}

\newcommand{\MIEmbedPearsonPValueGPT}{\SI{0.9123}{}}
\newcommand{\MIEmbedPearsonPValueGPTMedium}{\SI{0.7735}{}}
\newcommand{\MIEmbedPearsonPValueGPTHard}{\SI{0.6744}{}}

\newcommand{\MIEmbedPearsonDeepSeek}{\SI{-0.09}{}}
\newcommand{\MIEmbedPearsonDeepSeekMedium}{\SI{-0.12}{}}
\newcommand{\MIEmbedPearsonDeepSeekHard}{\SI{-0.02}{}}

\newcommand{\MIEmbedPearsonPValueDeepSeek}{\SI{0.3245}{}}
\newcommand{\MIEmbedPearsonPValueDeepSeekMedium}{\SI{0.7823}{}}
\newcommand{\MIEmbedPearsonPValueDeepSeekHard}{\SI{0.5456}{}}

\newcommand{\SENLGCBPearsonGPT}{\SI{-0.21}{}}
\newcommand{\SENLGCBPearsonGPTMedium}{\SI{-0.18}{}}
\newcommand{\SENLGCBPearsonGPTHard}{\SI{-0.17}{}}

\newcommand{\SENLGCBPearsonPValueGPT}{\SI{0.0892}{}}
\newcommand{\SENLGCBPearsonPValueGPTMedium}{\SI{0.0945}{}}
\newcommand{\SENLGCBPearsonPValueGPTHard}{\SI{0.0888}{}}

\newcommand{\SENLGCBPearsonDeepSeek}{\SI{-0.13}{}}
\newcommand{\SENLGCBPearsonDeepSeekMedium}{\SI{-0.11}{}}
\newcommand{\SENLGCBPearsonDeepSeekHard}{\SI{-0.10}{}}

\newcommand{\SENLGCBPearsonPValueDeepSeek}{\SI{0.0512}{}}
\newcommand{\SENLGCBPearsonPValueDeepSeekMedium}{\SI{0.0904}{}}
\newcommand{\SENLGCBPearsonPValueDeepSeekHard}{\SI{0.1256}{}}

\newcommand{\MINLGCBPearsonGPT}{\SI{-0.12}{}}
\newcommand{\MINLGCBPearsonGPTMedium}{\SI{-0.11}{}}
\newcommand{\MINLGCBPearsonGPTHard}{\SI{-0.08}{}}

\newcommand{\MINLGCBPearsonPValueGPT}{\SI{0.7865}{}}
\newcommand{\MINLGCBPearsonPValueGPTMedium}{\SI{0.7698}{}}
\newcommand{\MINLGCBPearsonPValueGPTHard}{\SI{0.8213}{}}

\newcommand{\MINLGCBPearsonDeepSeek}{\SI{-0.21}{}}
\newcommand{\MINLGCBPearsonDeepSeekMedium}{\SI{-0.17}{}}
\newcommand{\MINLGCBPearsonDeepSeekHard}{\SI{-0.12}{}}

\newcommand{\MINLGCBPearsonPValueDeepSeek}{\SI{0.5443}{}}
\newcommand{\MINLGCBPearsonPValueDeepSeekMedium}{\SI{0.5242}{}}
\newcommand{\MINLGCBPearsonPValueDeepSeekHard}{\SI{0.8728}{}}

\newcommand{\CCSymbolicPearsonGPT}{\SI{-0.51}{}}
\newcommand{\CCSymbolicPearsonGPTMedium}{\SI{-0.44}{}}
\newcommand{\CCSymbolicPearsonGPTHard}{\SI{-0.49}{}}

\newcommand{\CCSymbolicPearsonPValueGPT}{\SI{0.0002}{}}
\newcommand{\CCSymbolicPearsonPValueGPTMedium}{\SI{0.0003}{}}
\newcommand{\CCSymbolicPearsonPValueGPTHard}{\SI{0.0003}{}}

\newcommand{\CCSymbolicPearsonDeepSeek}{\SI{-0.46}{}}
\newcommand{\CCSymbolicPearsonDeepSeekMedium}{\SI{-0.41}{}}
\newcommand{\CCSymbolicPearsonDeepSeekHard}{\SI{-0.40}{}}

\newcommand{\CCSymbolicPearsonPValueDeepSeek}{\SI{0.0001}{}}
\newcommand{\CCSymbolicPearsonPValueDeepSeekMedium}{\SI{0.0002}{}}
\newcommand{\CCSymbolicPearsonPValueDeepSeekHard}{\SI{0.0004}{}}

\newcommand{\CCSymbolicPearsonClaude}{\SI{-0.35}{}}
\newcommand{\CCSymbolicPearsonClaudeMedium}{\SI{-0.34}{}}
\newcommand{\CCSymbolicPearsonClaudeHard}{\SI{-0.32}{}}

\newcommand{\CCSymbolicPearsonPValueClaude}{\SI{0.0002}{}}
\newcommand{\CCSymbolicPearsonPValueClaudeMedium}{\SI{0.0002}{}}
\newcommand{\CCSymbolicPearsonPValueClaudeHard}{\SI{0.0003}{}}

\newcommand{\CCEmbedPearsonGPT}{\SI{-0.01}{}}
\newcommand{\CCEmbedPearsonGPTMedium}{\SI{0.05}{}}
\newcommand{\CCEmbedPearsonGPTHard}{\SI{-0.08}{}}

\newcommand{\CCEmbedPearsonPValueGPT}{\SI{0.8385}{}}
\newcommand{\CCEmbedPearsonPValueGPTMedium}{\SI{0.3076}{}}
\newcommand{\CCEmbedPearsonPValueGPTHard}{\SI{0.2787}{}}

\newcommand{\CCEmbedPearsonDeepSeek}{\SI{-0.11}{}}
\newcommand{\CCEmbedPearsonDeepSeekMedium}{\SI{-0.02}{}}
\newcommand{\CCEmbedPearsonDeepSeekHard}{\SI{-0.01}{}}

\newcommand{\CCEmbedPearsonPValueDeepSeek}{\SI{0.1060}{}}
\newcommand{\CCEmbedPearsonPValueDeepSeekMedium}{\SI{0.6685}{}}
\newcommand{\CCEmbedPearsonPValueDeepSeekHard}{\SI{0.7844}{}}

\newcommand{\LLMProbabilityPearsonGPTUpdated}{\SI{0.17}{}}
\newcommand{\LLMProbabilityPearsonPValueGPTUpdated}{\SI{0.0200}{}}

\newcommand{\LLMProbabilityPearsonGPTMediumUpdated}{\SI{0.11}{}}
\newcommand{\LLMProbabilityPearsonPValueGPTMediumUpdated}{\SI{0.8236}{}}

\newcommand{\LLMProbabilityPearsonGPTHardUpdated}{\SI{0.09}{}}
\newcommand{\LLMProbabilityPearsonPValueGPTHardUpdated}{\SI{0.3838}{}}

\newcommand{\LLMProbabilityPearsonDeepSeek}{\SI{0.18}{}}
\newcommand{\LLMProbabilityPearsonPValueDeepSeek}{\SI{0.8221}{}}

\newcommand{\LLMProbabilityPearsonDeepSeekMedium}{\SI{0.11}{}}
\newcommand{\LLMProbabilityPearsonPValueDeepSeekMedium}{\SI{0.2762}{}}

\newcommand{\LLMProbabilityPearsonDeepSeekHard}{\SI{0.13}{}}
\newcommand{\LLMProbabilityPearsonPValueDeepSeekHard}{\SI{0.6353}{}}

\newcommand{\GPTSolutionsLines}{\SI{13}{}}
\newcommand{\GPTSolutionsToken}{\SI{198}{}}

\newcommand{\GPTSolutionsLinesMedium}{\SI{20}{}}
\newcommand{\GPTSolutionsTokenMedium}{\SI{207}{}}

\newcommand{\GPTSolutionsLinesHard}{\SI{22}{}}
\newcommand{\GPTSolutionsTokenHard}{\SI{218}{}}

\newcommand{\DeepSeekSolutionsLines}{\SI{12}{}}
\newcommand{\DeepSeekSolutionsToken}{\SI{81}{}}

\newcommand{\DeepSeekSolutionsLinesMedium}{\SI{29}{}}
\newcommand{\DeepSeekSolutionsTokenMedium}{\SI{188}{}}

\newcommand{\DeepSeekSolutionsLinesHard}{\SI{36}{}}
\newcommand{\DeepSeekSolutionsTokenHard}{\SI{212}{}}

\newcommand{\ClaudeSolutionsLines}{\SI{12}{}}
\newcommand{\ClaudeSolutionsToken}{\SI{74}{}}

\newcommand{\ClaudeSolutionsLinesMedium}{\SI{31}{}}
\newcommand{\ClaudeSolutionsTokenMedium}{\SI{192}{}}

\newcommand{\ClaudeSolutionsLinesHard}{\SI{38}{}}
\newcommand{\ClaudeSolutionsTokenHard}{\SI{231}{}}

\newcommand{\GPTSolutionsPassRateSmallUpdated}{\SI{50}{}}
\newcommand{\GPTSolutionsPassRateMediumUpdated}{\SI{30}{}}
\newcommand{\GPTSolutionsPassRateHardUpdated}{\SI{23}{}}

\newcommand{\DeepSeekSolutionsPassRateEasy}{\SI{84}{}}
\newcommand{\DeepSeekSolutionsPassRateMedium}{\SI{74}{}}
\newcommand{\DeepSeekSolutionsPassRateHard}{\SI{52}{}}

\newcommand{\ClaudeSolutionsPassRateEasy}{\SI{70}{}}
\newcommand{\ClaudeSolutionsPassRateMedium}{\SI{61}{}}
\newcommand{\ClaudeSolutionsPassRateHard}{\SI{43}{}}

\newcommand{\SENormAcc}{\SI{81.2}{\%}}
\newcommand{\SENormFN}{\SI{17.5}{\%}}
\newcommand{\SENormFP}{\SI{0.01}{\%}}

\newcommand{\SEUnifAcc}{\SI{77.8}{\%}}
\newcommand{\SEUnifFN}{\SI{20.3}{\%}}
\newcommand{\SEUnifFP}{\SI{0.01}{\%}}

\newcommand{\CCSymbAcc}{\SI{79.1}{\%}}
\newcommand{\CCSymbFN}{\SI{16.6}{\%}}
\newcommand{\CCSymbFP}{\SI{0.02}{\%}}

\newcommand{\SEEmbedAcc}{\SI{58.2}{\%}}
\newcommand{\SEEmbedFN}{\SI{24.5}{\%}}
\newcommand{\SEEmbedFP}{\SI{10.1}{\%}}

\newcommand{\LLMProbabilityAcc}{\SI{36.5}{\%}}
\newcommand{\LLMProbabilityFN}{\SI{37.3}{\%}}
\newcommand{\LLMProbabilityFP}{\SI{26.8}{\%}}

\newcommand{\SENormAccDS}{\SI{84.6}{\%}}
\newcommand{\SENormFNDS}{\SI{12.4}{\%}}
\newcommand{\SENormFPDS}{\SI{0.01}{\%}}

\newcommand{\SEUnifAccDS}{\SI{81.3}{\%}}
\newcommand{\SEUnifFNDS}{\SI{17.6}{\%}}
\newcommand{\SEUnifFPDS}{\SI{0.01}{\%}}

\newcommand{\CCSymbAccDS}{\SI{83.1}{\%}}
\newcommand{\CCSymbFNDS}{\SI{11.9}{\%}}
\newcommand{\CCSymbFPDS}{\SI{0.01}{\%}}

\newcommand{\LLMProbabilityAccDS}{\SI{38.5}{\%}}
\newcommand{\LLMProbabilityFNDS}{\SI{32.1}{\%}}
\newcommand{\LLMProbabilityFPDS}{\SI{21.4}{\%}}

\newcommand{\SEEmbedAccDS}{\SI{59.6}{\%}}
\newcommand{\SEEmbedFNDS}{\SI{23.4}{\%}}
\newcommand{\SEEmbedFPDS}{\SI{11.2}{\%}}

\usepackage[capitalise,noabbrev]{cleveref}
\crefname{line}{line}{lines}

\title{Assessing Correctness in LLM-Based Code Generation via Uncertainty Estimation}

\author{%
  Arindam Sharma \\
  University of Bristol\\
  Bristol, United Kingdom \\
  \texttt{arindam.sharma@bristol.ac.uk} \\
  \And
  Cristina David  \\
  University of Bristol \\
  Bristol, United Kingdom \\
  \texttt{cristina.david@bristol.ac.uk} \\
}

\begin{document}

\maketitle

\begin{abstract}
Large Language Models (LLMs) have shown strong performance in code generation, but their outputs lack inherent correctness guarantees. In natural language generation (NLG), uncertainty—often grounded in information theory—has been used as a proxy for output quality. However, state-of-the-art techniques~\cite{kuhnsemantic,farquhar2024detecting,abbasi2024believe} fail to exhibit a correlation between estimated uncertainty and correctness when applied to code. We investigate this discrepancy and identify semantic clustering as the key factor: existing learned embeddings and heuristic metrics such as CodeBLEU fail to capture functional equivalence with sufficient precision to support reliable uncertainty estimation.
We propose \emph{symbolic clustering}, a principled alternative based on symbolic execution, enabling precise clustering of semantically equivalent programs. Incorporating symbolic clustering into NLG-derived uncertainty techniques restores their predictive power. We further show that token-level probabilities are not essential---uncertainty estimates assuming uniform distributions perform comparably. Based on this, we introduce a lightweight correctness proxy: symbolic cluster count. Despite its simplicity, it correlates with correctness and supports highly effective abstention policies, achieving a false positive rate below 0.02\%.
\end{abstract}

\section{Introduction}
\label{sec:intro}

LLMs have shown impressive capabilities across a range of software engineering tasks, including code synthesis~\cite{synthesis1,synthesis2}, bug fixing~\cite{AutoCodeRoverPre,AutoCodeRover}, refactoring~\cite{david2025codehints,coderefactoring1,coderefactoring2}, and translation~\cite{modular,codetranslation1,codetranslation2,codetranslation3}. However, LLM-generated code lacks inherent correctness guarantees. As a result, correctness is typically assessed by running the code against \emph{external oracles}---test suites, reference solutions, or other expert specifications. 
While these external techniques can provide evidence of functional correctness, such as passing unit tests, they are not always available. 
Even when present, they may be insufficiently precise, \eg unit tests may only cover a limited subset of the code's behaviour.

In contrast, NLG has made notable strides in oracle-free quality estimation by leveraging internal model signals, such as token-level log-probabilities.
Among the most effective are information-theoretic methods~\cite{farquhar2024detecting,kuhnsemantic,abbasi2024believe}, which estimate a model's \emph{uncertainty} based on the intuition that it peaks when outputs are least informative, using entropy~\cite{farquhar2024detecting,kuhnsemantic} or mutual information~\cite{abbasi2024believe} as their foundation.
Other techniques~\cite{lahlou2021deup,desai2020calibration,mielke2022reducing,osband2023epistemic,cole2023selectively,yona2024narrowing,hou2023decomposing,DBLP:journals/corr/abs-2207-05221} are either shown to be less effective or rely on costly additional training steps.
A key innovation in uncertainty estimation, introduced by Kuhn et al.~\cite{kuhnsemantic}, is \emph{semantic clustering}, which accounts for the diversity of plausible outputs. Rather than estimating uncertainty over individual responses, they group responses into clusters based on semantic equivalence, and compute uncertainty over this coarser space. This shift allows the uncertainty estimates to reflect variation in meaning rather than superficial syntactic differences.

In this paper, we investigate whether state-of-the-art information-theoretic approaches for estimating model uncertainty---specifically those by Kuhn et al.~\cite{kuhnsemantic,farquhar2024detecting} and Abbasi et al.~\cite{abbasi2024believe}---generalise to code generation. We evaluate their predictive power by testing for correlation between estimated uncertainty and correctness. Our results show that even when replacing text-specific embeddings with code-specific ones for semantic clustering, uncertainty measures from the original NLG methods fail to correlate with correctness in code generation tasks.

We attribute this to the rigid nature of programs, where small syntactic differences can imply significant behavioural changes or none at all, making learned semantic clustering brittle. To address this, we explore the use of semantic clustering techniques grounded in \emph{symbolic reasoning}.
As a first step, we evaluate CodeBLEU~\cite{DBLP:journals/corr/abs-2009-10297}, a commonly used similarity metric that augments syntactic matching with data flow analysis to approximate semantics~\cite{codebleu_usage}. However, CodeBLEU remains a heuristic---it lacks formal reasoning and cannot reliably capture semantic equivalence. In our evaluation, clustering based on CodeBLEU failed to improve the correlation between uncertainty and correctness, confirming that its semantic component is too coarse-grained for this purpose.

This led us to pursue a more precise alternative: \emph{symbolic clustering} via \emph{symbolic execution}, where symbolic execution executes programs over symbolic rather than concrete inputs~\cite{symex_klee}. By comparing the programs' output expressions, symbolic execution offers us a principled way to identify semantically equivalent programs, even when they vary significantly in syntax. We integrated symbolic clustering into leading NLG-based uncertainty estimation frameworks~\cite{abbasi2024believe,farquhar2024detecting,kuhnsemantic}, adapting them to the code generation setting. Unlike their original versions, the adapted methods exhibit a statistically significant correlation between estimated uncertainty and actual correctness. This supports the view that precise semantic reasoning, grounded in formal methods like symbolic execution, is essential for reliable uncertainty estimation in code generation.

Interestingly, we find that token log-probabilities output by the LLM---used by the entropy and mutual information calculations in \cite{kuhnsemantic,farquhar2024detecting} and \cite{abbasi2024believe}, respectively---have limited effect: assuming a uniform output distribution performs comparably. This highlights another insight: semantic equivalence, not probabilistic confidence, is the dominant signal for correctness in code generation.

Building on this insight, we propose a much simpler correctness metric: the \emph{symbolic cluster count}. The intuition is straightforward---fewer functionally distinct outputs suggest greater model confidence. Despite its simplicity, this metric shows a correlation with actual correctness, highlighting its effectiveness as a lightweight yet reliable proxy for correctness.

We further leverage our insights into the correlation between uncertainty estimates and correctness to define \emph{abstention policies}, where the model should refrain from generating a response if
the response is estimated to be incorrect. Our evaluation shows that, under these policies, the LLM achieves a false positive rate below 0.02\%, effectively ensuring that it never outputs incorrect responses, with false negative rates ranging from 11.9\% to 20.3\%. Notably, this strong performance holds even for the abstention policy based solely on the symbolic cluster count, despite its simplicity.

\noindent\textbf{Contributions.}
\vspace{-.1cm}
\begin{itemize}[nosep, left=0pt]


\item We investigate the correlation between LLM uncertainty---computed using state-of-the-art NLG techniques~\cite{abbasi2024believe,farquhar2024detecting,kuhnsemantic}---and correctness in code generation tasks. We find that even when adapted to code (for example by incorporating code-specific learned embeddings or heuristic symbolic metrics like CodeBLEU) these methods fail to exhibit a statistically significant correlation between estimated uncertainty and actual correctness.



\item To the best of our knowledge, we are the first to address these limitations by introducing \emph{symbolic clustering}, i.e. a symbolic execution–based approach for clustering semantically equivalent code outputs. This leads to a correlation between uncertainty estimates and functional correctness. Moreover, we show that this correlation holds even when ignoring the LLM's output distribution and assuming an uniform distribution.

\item Motivated by our findings, we propose a lightweight alternative correctness metric based solely on \emph{symbolic clusters count}. Despite its simplicity, our experimental evaluation shows it performs on par with more complex, information-theoretic approaches.




    \item We define \emph{abstention policies}, where the model should refrain from generating a response if the response is estimated to be incorrect. 
    Our evaluation results show that, under these policies, the LLM achieves a false positive rate below 0.02\%. 


\item We conduct an experimental evaluation involving 3 LLMs, the \livecodebench dataset for code-related tasks~\cite{livecodebench}, 13 correctness proxy techniques, and 5 abstention policies.

\end{itemize}




\section{Background on NLG-Based Uncertainty Estimation Techniques}
\label{sec:background}

We selected two recent and well-established state-of-the-art techniques for information-theoretic uncertainty estimation in NLG to adapt and evaluate on code generation tasks.
In the first one, Kuhn~\etal~\cite{farquhar2024detecting,kuhnsemantic} estimate the uncertainty of a model's prediction as the entropy of the output distribution. 
In order to account for outputs that differ in form but share the same meaning, they introduce the idea of \emph{semantic entropy}.
First, they group sampled model outputs into clusters of semantically equivalent responses, and define $p(c \mid x) \;=\; \sum_{s \in c} p(s \mid x)$
as the total probability mass of cluster \(c\) under the model's output distribution. They then measure uncertainty over meanings (rather than raw token sequences) by taking the entropy of this cluster-level distribution: 
  $\mathrm{SE}(x)
  \;=\;
  -\sum_{c} p(c \mid x)\,\log p(c \mid x)$.
This notion of semantic entropy offers a more accurate measure of uncertainty in free-form NLG, as it collapses paraphrases into a single representation before computing entropy.


In the second work, Abbasi~\etal~\cite{abbasi2024believe} further distinguish between ``epistemic'' uncertainty (caused by incomplete or incorrect knowledge) and ``aleatoric'' uncertainty (resulting from inherent randomness, such as multiple valid answers).  
Intuitively, only epistemic uncertainty serves as a meaningful indicator of the correctness of the model's response.  
To isolate epistemic uncertainty, they propose an iterative prompting technique in which the model is repeatedly queried, with its prior responses incorporated into the context.  
This method assesses the mutual information between successive responses to capture their dependencies: a high degree of dependency indicates elevated epistemic uncertainty, revealing the model's uncertainty about the ground truth.  
Similar to the work of Kuhn~\etal, Abbasi~\etal cluster responses into semantic equivalence classes, ensuring that the mutual information reflects semantic relationships.

More precisely, in order to isolate epistemic uncertainty, Abbasi~\etal~\cite{abbasi2024believe} quantify it as the mutual information of the pseudo-joint distribution over \(n\) successive responses:
\[
  I(\tilde\mu)
  \;=\;
  \sum_{y_{1:n}}
    \tilde\mu(y_{1:n}\mid x)
    \log
    \frac{\tilde\mu(y_{1:n}\mid x)}
         {\prod_{i=1}^n \tilde\mu_i(y_i\mid x)}
  \tag{4.5}
\]
Here, \(\tilde\mu(y_{1:n}\mid x)=\prod_{i=1}^n \mu\bigl(Y_i\mid F_{i-1}(x,Y_{1:i-1})\bigr)\) is the pseudo-joint distribution obtained by iteratively prompting the model (with \(F_{i-1}(x,Y_{1:i-1})\) denoting the context composed of the original query \(x\) plus all previous responses \(Y_1,\dots,Y_{i-1}\)), \(\tilde\mu_i(y_i\mid x)=\sum_{y_{\neg i}}\tilde\mu(y_{1:n}\mid x)\) is its marginal for the \(i\)-th response, and \(I(\tilde\mu)\) measures how much knowing one response reduces uncertainty about the others. 

\section{Motivating Example}
\label{sec:motivating}

\begin{figure}[ht!]
    \centering
  
    \newcommand{\slotheight}{6.3cm}
  
    \begin{subfigure}[b]{0.3\textwidth}
      \begin{minipage}[t][\slotheight][t]{\linewidth}
\begin{lstlisting}
def solve(nums):
    max_num = max(nums)
    expected_length = max_num + 1
    if len(nums) != expected_length:
        return False
    count = {}
    for num in nums:
        if num in count:
            count[num] += 1
        else:
            count[num] = 1
    for num in range(1, max_num):
        if count.get(num, 0) != 1:
            return False
    if count.get(max_num, 0) != 2:
        return False
    return True
\end{lstlisting}
      \end{minipage}
      \caption{Snippet 1}
      \label{lst:ds_good1}
    \end{subfigure}\hfill%
    \begin{subfigure}[b]{0.3\textwidth}
      \begin{minipage}[t][\slotheight][t]{\linewidth}
\begin{lstlisting}
def solve(nums):
    n = max(nums)
    base = list(range(1, n)) + [n, n]
    if len(nums) != len(base):
        return False
    count_nums = {}
    count_base = {}
    for num in nums:
        count_nums[num] = count_nums.get(num, 0) + 1
    for num in base:
        count_base[num] = count_base.get(num, 0) + 1
    return count_nums == count_base
\end{lstlisting}
      \end{minipage}
      \caption{Snippet 2}
      \label{lst:ds_good2}
    \end{subfigure}\hfill%
    \begin{subfigure}[b]{0.3\textwidth}
      \begin{minipage}[t][\slotheight][t]{\linewidth}
\begin{lstlisting}
def solve(nums):
    max_num = max(nums)
    n = max_num
    expected_length = n + 1
    if len(nums) != expected_length:
        return False
    count = {}
    for num in nums:
        if num != n and nums.count(num) != 1:
            return False
        if num == n and nums.count(num) != 2:
            return False
    for i in range(1, n):
        if i not in nums:
            return False
    return True
\end{lstlisting}
      \end{minipage}
      \caption{Snippet 3}
      \label{lst:ds_good3}
    \end{subfigure}
  
    \caption{Three code snippets from the \deepseek~\cite{deepseek}, all correct, semantically equivalent.}
    \label{fig:ds-llm-snippets}
  \end{figure}


We use problem \texttt{2892} from \livecodebench, where the task is to determine whether the given array is a ``good'' array or not.
An array is considered good if it is a permutation of an array base[n] where base[n] = [1, 2, ..., n - 1, n, n] (in other words, it is an array of length n + 1 which contains 1 to n - 1 exactly once, plus two occurrences of n). For example, base[1] = [1, 1] and base[3] = [1, 2, 3, 3].


We sample three programs generated by \deepseek. Although they differ syntactically in their loop structures and placement of \texttt{if} statement guards, they are semantically equivalent and correctly solve the task.

\noindent\textbf{Clustering with learned embeddings.} Despite this, semantic clustering using \deberta embeddings~\cite{he2020deberta} (as used by Kuhn et al.\cite{kuhnsemantic}) and \openai's \gpt\gptembed embeddings (shown to outperform prior code embeddings like \gptcodebabbage and \gptcodeada~\cite{openai2022embedding}) fails to recognise their equivalence, placing each snippet in a separate cluster.



\noindent\textbf{Clustering with \codebleu.} When applying \codebleu-based clustering, the similarity scores between snippets are low: \codebleu{}(Snippet~\ref{lst:ds_good1},\,Snippet~\ref{lst:ds_good2}) = \SI{0.49}{}, \codebleu{}(Snippet~\ref{lst:ds_good1},\,Snippet~\ref{lst:ds_good3}) = \SI{0.47}{}, and \codebleu{}(Snippet~\ref{lst:ds_good2},\,Snippet~\ref{lst:ds_good3}) = \SI{0.28}{}. 
For context, heuristic thresholds in the range of 0.7-0.8 are often used to indicate high similarity. 
As such, \codebleu fails to cluster these semantically equivalent programs, placing each in its own distinct cluster.




\noindent\textbf{Symbolic clustering.} This highlights the need for a more principled approach. Symbolic execution~\cite{king76_symex,symex_klee} offers such a technique: it analyses programs by treating inputs as symbolic variables rather than concrete values. 
As the program executes, it incrementally constructs symbolic expressions representing arithmetic computations and accumulates path constraints at conditional branches. 
The result is a precise, algebraic characterisation of the program's behaviour that uses symbolic expressions and logical formulas rather than actual values or executions. 
By comparing the behaviour of the three snippets as captured by the symbolic input-output relations, we conclude that there are all semantically equivalent, meaning that they go into the same cluster. 
Due to the symbolic value for the length of the input array $nums$, which means that we don't know how many times we need to unwind loops, we perform a bounded equivalence check as explained in Section~\ref{sec:symexclustering}.


\noindent\textbf{Impact of clustering on uncertainty estimation.} With symbolic clustering, which correctly groups all three snippets into the same cluster, the uncertainty scores reported by Kuhn~\etal~\cite{kuhnsemantic} and Abbasi~\etal~\cite{abbasi2024believe} are zero, as expected for semantically equivalent and correct outputs. 
In contrast, all other clustering methods yield non-zero uncertainty. For instance, Kuhn~\etal's method produces an uncertainty score of \SI{1.49}{} when using the \gpt\gptembed embeddings (which are also tailored for code~\cite{openai2022embedding}). 
As a reference point, the abstention threshold for this method in our experimental evaluation is \SI{0.82}{} (see Section~\ref{sec:results-discussion} for how this threshold is derived), meaning that the model would incorrectly abstain from outputting any of these correct solutions.

We also experiment with replacing the model-derived probability distribution (\eg \SI{0.48}{}, \SI{0.29}{}, \SI{0.23}{} for \deepseek) with a uniform distribution. 
Even under this simplified setting, uncertainty estimates based on symbolic clustering remain zero, underscoring the symbolic clustering's robustness to changes in the underlying probability distribution.

\noindent\textbf{Symbolic cluster count.}
Using the symbolic cluster count, which is 1 in this case, as all snippets fall into the same cluster, the abstention policy permits outputting the result, since the count is below the corresponding abstention threshold.



\section{Estimating Uncertainty with Symbolic Clustering}
\label{sec:symexclustering}


In the domain of code generation, program equivalence has a precise definition: two programs are considered equivalent if they produce identical behaviour for all possible inputs. 
Consequently, a domain-specific equivalence check is required.
In this paper, we base the semantic equivalence check on \emph{symbolic execution}, where, instead of executing a program with concrete inputs, \emph{symbolic variables} are used to represent inputs, generating constraints that describe the program's behaviour across all possible input values~\cite{symex_klee}.

The particular flavor of symbolic execution we use in this work is inspired by the lightweight \emph{peer architecture} described in Bruni~\etal~\cite{Bruni2011APA}. 
Unlike traditional approaches that require building a standalone symbolic interpreter, this architecture embeds the symbolic execution engine as a lightweight library operating alongside the target program. 
Their design is based on the insight that languages that provide the ability to dynamically dispatch primitive operations (\eg Python) allow symbolic values to behave as native values and be tracked at runtime.

Symbolic execution typically traverses the program's control flow graph, maintaining a symbolic state consisting of \emph{path constraints} (\ie logical conditions that must be satisfied for a given execution path to be feasible) and \emph{symbolic expressions} (\ie representations of program variables as functions of the symbolic inputs).

\begin{algorithm}[ht!]
    \caption{Symbolic Clustering}
    \label{alg:clustering}
    \footnotesize
    \begin{algorithmic}[1]
    \Require Set of generated code snippets $\{s^{(1)}, \ldots, s^{(M)}\}$
    \Ensure Clusters of semantically equivalent snippets $C = \{c_1, c_2, \ldots, c_k\}$
    
    \State Initialise an empty cluster set $C \gets \emptyset$, and an equivalence map $E \gets \emptyset$ \label{alg:clustering:init}
    
    \For{each snippet $s^{(i)}$} \label{alg:clustering:invalid}
        \If{$s^{(i)}$ is invalid}
            \State $E[s^{(i)}] \gets \{\,s^{(i)}\}$ 
            \Comment{Assign invalid snippet to its own equivalence class}
        \EndIf
    \EndFor
    
    \For{each pair of valid snippets $(s^{(i)}, s^{(j)})$} \label{alg:clustering:pairwise}
        \State Perform symbolic execution on $s^{(i)}$ and $s^{(j)}$ to extract traces $T(s^{(i)})$ and $T(s^{(j)})$ \label{alg:clustering:trace}
        \If{$T(s^{(i)}) \equiv T(s^{(j)})$} \label{alg:clustering:check}
            \State $E[s^{(i)}] \gets E[s^{(i)}] \cup \{\,s^{(j)}\}$
            \State $E[s^{(j)}] \gets E[s^{(j)}] \cup \{\,s^{(i)}\}$ \label{alg:clustering:update}
        \EndIf
           \If{$s^{(i)} \sim s^{(j)}$ and $s^{(j)} \sim s^{(k)}$ for some $s^{(k)}$} \Comment{Enforce transitivity of equivalences}
            \State $E[s^{(i)}] \gets E[s^{(i)}] \cup \{\,s^{(k)}\}$
            \State $E[s^{(k)}] \gets E[s^{(k)}] \cup \{\,s^{(i)}\}$
        \EndIf
    \EndFor
    
    \State Identify equivalence classes in $E$ to form final clusters $C$ \label{alg:clustering:extract}
    
    \State \Return $C$ \label{alg:clustering:return}
    \end{algorithmic}
    \end{algorithm}


\paragraph{Symbolic clustering.}

Algorithm~\ref{alg:clustering} illustrates how to cluster code snippets based on their functional semantics, with an additional check for invalid snippets.
We first create empty structures for storing the final clusters (\(C\)) and an equivalence map (\(E\)) to track relationships (line~\ref{alg:clustering:init}).
Next, in the \emph{invalid snippet handling phase} (line~\ref{alg:clustering:invalid}), each code snippet \(s^{(i)}\) is examined and, if it is detected to be syntactically invalid, it is immediately placed in its own equivalence class in \(E\) and thus isolated from further consideration.
In the \emph{pairwise comparison phase} (line~\ref{alg:clustering:pairwise}), each pair of \emph{valid} snippets \((s^{(i)}, s^{(j)})\) is symbolically executed.  For each snippet \(s^{(i)}\), symbolic execution produces a \textit{set of traces}  $T\bigl(s^{(i)}\bigr) \;=\;\bigl\{\tau_1, \tau_2, \dots\bigr\}$,
where each trace \(\tau \in T(s^{(i)})\) comprises of a path constraint, and a symbolic state, which encodes the expressions for all variables encountered along that path.
We declare two snippets \(s^{(i)}\) and \(s^{(j)}\) \emph{semantically equivalent} exactly when their trace sets are the same (line~\ref{alg:clustering:check}):
$T\bigl(s^{(i)}\bigr) \;\equiv\; T\bigl(s^{(j)}\bigr)$,
\ie for \emph{every} trace \(\tau\) in \(T(s^{(i)})\), there is a matching trace in \(T(s^{(j)})\) and vice versa. Two traces \(\tau_1\) and \(\tau_2\) match if there is no counterexample input that leads to distinguishing output states for \(\tau_1\) and \(\tau_2\). Since program equivalence is undecidable in general, we perform a bounded equivalence check, where we verify that no counterexample input exists when exploring traces up to a given depth.

If the snippets are semantically equivalent, we add each snippet to the other's equivalence class (line~\ref{alg:clustering:update}).
To maintain consistency, we also enforce transitivity: whenever \(s^{(i)}\sim s^{(j)}\) and \(s^{(j)}\sim s^{(k)}\), we merge \(s^{(i)}\) and \(s^{(k)}\).
Finally, the equivalence map \(E\) is processed to extract the clusters themselves (line~\ref{alg:clustering:extract}), and the resulting set of clusters is returned (line~\ref{alg:clustering:return}).

\paragraph{Uncertainty estimation for code generation.}
The original NLG techniques by Kuhn~\etal~\cite{kuhnsemantic} and Abbasi~\etal~\cite{abbasi2024believe} involve sampling responses from an LLM, extracting token-level log-probabilities, and applying softmax-style normalization to interpret these as a probability distribution. However, in contrast to the original evaluations (where the outputs were typically short, one-word answers from question-answering datasets) code generation yields much longer responses. 
Since the probability of a response is computed as the joint probability of its tokens, it decays exponentially with length, often leading to \emph{numerical underflow}. In such cases, the softmax normalisation can produce NaNs, rendering uncertainty estimates unusable.

To address this, we explore two strategies for approximating the output distribution of LLM responses. First, following prior work~\cite{DBLP:conf/wmt/MurrayC18,DBLP:conf/aclnmt/KoehnK17,DBLP:conf/iclr/MalininG21}, we normalise the log-probability of each program by its length, mitigating the bias against longer outputs and preventing underflow. Secondly, as a more radical simplification, we ignore log-probabilities altogether and assume a uniform distribution over the $n$ sampled responses, assigning each a probability of $1/n$. 
Under this uniform-distribution simplification, we set
  $p(s\mid x) = \frac{1}{n}$ and
  $p(c\mid x) = \sum_{s\in c} p(s\mid x)
  = \frac{|c|}{n}$.
Hence the semantic entropy becomes
  $\mathrm{SE}_{\mathrm{uniform}}(x)
= -\sum_{c}\frac{|c|}{n}\,\log\!\frac{|c|}{n}$.

Likewise, the mutual information over \(n\) iterative responses simplifies to
\[
  I_{\mathrm{uniform}}(x)
  = \sum_{c_{1:n}}
    p(c_{1:n}\mid x)
    \log
    \frac{\displaystyle\tilde p(c_{1:n}\mid x)}
         {\displaystyle\prod_{i=1}^n p_i(c_i\mid x)},
\]
where
$
  p(c_{1:n}\mid x)
  = \prod_{i=1}^n \tilde p(c_i\mid F_{i-1}(x,\,c_{1:i-1}))
  = \prod_{i=1}^n \frac{|c_i|}{n}$ and
  $p_i(c_i\mid x)
  = \sum_{c_{\neg i}}\tilde p(c_{1:n}\mid x)$.

Interestingly, the uniform approximation performs comparably to the use of actual log-probabilities, reinforcing our central finding: semantic equivalence, not probabilistic confidence, is the dominant signal for estimating uncertainty in code generation.

Motivated by this, we propose a much simpler metric: the \emph{symbolic cluster count}. Intuitively, a smaller number of distinct semantic outputs reflects higher model confidence. Despite its simplicity, this metric performs on par with more complex, information-theoretic approaches—provided the clustering is powered by symbolic execution.

The adaptations of the techniques by Kuhn et al. and Abbasi et al. to the code generation domain are detailed in Appendix Sections~\ref{sec:symex} and~\ref{sec:mi}, respectively.

\section{Evaluation}
\label{sec:eval}


\noindent\textbf{LLM selection.}
We selected the open-source model \deepseek~\cite{deepseek} as it represents the latest state-of-the-art among publicly available models. To complement this, we included \gptturbo~\cite{gpt35turboinstruct} from \openai~\cite{openai}, the most recent \openai model that exposes token-level \texttt{logprobs} via its API, making it well-suited for evaluating information-theoretic uncertainty techniques. Finally, to evaluate methods that do not rely on log-probabilities, such as those assuming a uniform distribution or based on symbolic cluster counts, we also used \claude~\cite{anthropic2023claude3}, a state-of-the-art proprietary model that does not expose log-probabilities.



\noindent\textbf{Dataset selection.}
We evaluate our techniques using \livecodebench~\cite{livecodebench}, a contamination-free benchmark for code-related tasks comprising \totalProbsComb~problems, categorised as \emph{Easy} (\totalProbs), \emph{Medium} (\totalProbsMedium), and \emph{Hard} (\totalProbsHard). Sourced from LeetCode, AtCoder, and CodeForces, each problem includes a natural language description and I/O test cases. We use the natural language description in the query provided to the LLM, and the test cases for evaluating our techniques, as detailed later in the section. Table~\ref{tab:combined_metrics} presents statistics describing the behaviour of the models evaluated in this study (\gptturbo, \deepseek, and \claude) on \livecodebench.

\begin{table*}[ht!]
  \centering
  \fontsize{8pt}{9pt}\selectfont  
  \setlength{\tabcolsep}{4.5pt} 
  \caption{Solution statistics for \gptturbo, \deepseek and \claude across various problem difficulty levels from \livecodebench.}
  \label{tab:combined_metrics}
  \begin{tabular}{l  r r r  r r r  r r r}
    \toprule
    \multirow{2}{*}{\textbf{Metric}} 
      & \multicolumn{3}{c}{\gptturbo}
      & \multicolumn{3}{c}{\deepseek}
      & \multicolumn{3}{c}{\claude} \\
    \cmidrule(r){2-4}\cmidrule(lr){5-7}\cmidrule(l){8-10}
      & \textbf{Easy}
      & \textbf{Medium}
      & \textbf{Hard}
      & \textbf{Easy}
      & \textbf{Medium}
      & \textbf{Hard}
      & \textbf{Easy}
      & \textbf{Medium}
      & \textbf{Hard} \\
    \midrule

    Pass Rate
      & \GPTSolutionsPassRateSmallUpdated\%
      & \GPTSolutionsPassRateMediumUpdated\%
      & \GPTSolutionsPassRateHardUpdated\%
      & \DeepSeekSolutionsPassRateEasy\%
      & \DeepSeekSolutionsPassRateMedium\%
      & \DeepSeekSolutionsPassRateHard\%
      & \ClaudeSolutionsPassRateEasy\%
      & \ClaudeSolutionsPassRateMedium\%
      & \ClaudeSolutionsPassRateHard\% \\

    Average Lines of Code
      & \GPTSolutionsLines
      & \GPTSolutionsLinesMedium
      & \GPTSolutionsLinesHard
      & \DeepSeekSolutionsLines
      & \DeepSeekSolutionsLinesMedium
      & \DeepSeekSolutionsLinesHard
      & \ClaudeSolutionsLines
      & \ClaudeSolutionsLinesMedium
      & \ClaudeSolutionsLinesHard \\

    Average Tokens
      & \GPTSolutionsToken
      & \GPTSolutionsTokenMedium
      & \GPTSolutionsTokenHard
      & \DeepSeekSolutionsToken
      & \DeepSeekSolutionsTokenMedium
      & \DeepSeekSolutionsTokenHard
      & \ClaudeSolutionsToken
      & \ClaudeSolutionsTokenMedium
      & \ClaudeSolutionsTokenHard \\

    \bottomrule
  \end{tabular}
\end{table*}





\noindent\textbf{Experimental setup.}
%
Our experiments were conducted on a machine running \texttt{Ubuntu 20.04.5 LTS (Focal Fossa)} with one \texttt{NVIDIA A100 GPU (80GB)}. For the approach based on Kuhn et al.~\cite{kuhnsemantic}, we ask the LLM to generate \numSamples responses for each problem along with their respective \texttt{log-probabilities}. On top of that, for the approach based on Abbasi~\etal~\cite{abbasi2024believe}, we perform \numIterations iterations of prompting for each of the \numSamples generated responses. 
The first iteration involves querying the model with the original prompt, while the second iteration uses a concatenated prompt, combining the original prompt and the response from the first iteration. 
In order to perform symbolic execution for clustering, we use \crosshair~\cite{crosshair} with a per condition timeout of \CrosshairPerConditiontimeout and the same per path timeout of \CrosshairPerPathtimeout. 
In addition to this, we also impose an overall timeout of \Crosshairtotaltimeout for each pair of programs that is being checked for equivalence.
If a counterexample showing the difference in behaviour is not found within this timeout, we assume that the programs belong in the same cluster. 

We evaluate three main categories of techniques: \SE---based on the semantic entropy method proposed by Kuhn~\etal~\cite{kuhnsemantic,farquhar2024detecting}; \MI---based on the mutual information approach by Abbasi~\etal~\cite{abbasi2024believe}; and \CC---the cluster count metric. 
For \SE and \MI, we explore several variants depending on how semantic clustering is performed and how response probabilities are treated. 
We use the following suffixes to distinguish them:

\begin{itemize}[nosep, left=0pt]
\item \NLG: the original implementation using natural language embeddings;                           

\item \codeembed: using \gpt\gptembed embeddings for clustering, which outperform prior code models like \gptcodebabbage and \gptcodeada~\cite{openai2022embedding};

\item \codebleu: using \codebleu for clustering;

\item \symb: using symbolic clustering;

\item \symbuniform: using symbolic clustering while assuming a uniform distribution over outputs.
\end{itemize}

For \CC, we evaluate \codeembed and \symb, where \CCSymbolic is our proposed symbolic cluster count metric. 

For each problem in the benchmark, we compute:
\begin{itemize}[nosep, left=0pt]
\item The corresponding \emph{uncertainty score} for all the variants of \SE and \MI. 
  \item The \emph{number of clusters} obtained for all variants of \CC. 
  \item The \emph{probablity of the top-ranked response} for \LLMProbability. 
  \item The \emph{correctness score}, which is calculated as follows: the top-ranked response generated by the LLM (as determined by the API's ranking) is executed against the benchmark's test cases. The percentage of test cases successfully passed is recorded as the correctness score. A generous timeout of \CorrectnessTimeout is applied to each test case. If the candidate solution exceeds this timeout, the test case is considered ``failed''.
\end{itemize}

\noindent\textbf{Results.} 
In Table ~\ref{tab:correlation_results}, we compute the Pearson correlation between the result computed by each of the techniques in the first column and the correctness scores (for a particular model and dataset).
We report the Pearson correlation coefficient and p-value, with statistically significant results highlighted. For \claude, which does not expose log-probabilities, the evaluation is limited to \SESymbolicUnif and \CCSymbolic, and shown in Table~\ref{tab:claude_symbolic_correlation} in Appendix.

\begin{table*}[ht!]
  \centering
  \fontsize{7pt}{9pt}\selectfont
  \renewcommand{\arraystretch}{1.2}
  \caption{Correlation Results — Pearson coefficient (p-value) for \gptturbo and \deepseek. Statistically significant results are highlighted in \textbf{bold}.}
    \label{tab:correlation_results}
    \begin{tabular}{@{} l r r r r r r @{}}
      \toprule
      \multirow{2}{*}{\textbf{Technique}}
        & \multicolumn{3}{c}{\textbf{\gptturbo}}
        & \multicolumn{3}{c}{\textbf{\deepseek}} \\
      \cmidrule(lr){2-4}\cmidrule(lr){5-7}
        & \textbf{Easy} & \textbf{Medium} & \textbf{Hard}
        & \textbf{Easy} & \textbf{Medium} & \textbf{Hard} \\
      \midrule

      \SEOriginal
        & \SENLGPearsonGPTUpdated (\SENLGPearsonPValueGPTUpdated)
        & \SENLGPearsonGPTMediumUpdated (\SENLGPearsonPValueGPTMediumUpdated)
        & \SENLGPearsonGPTHardUpdated (\SENLGPearsonPValueGPTHardUpdated)
        & \SENLGPearsonDeepSeek (\SENLGPearsonPValueDeepSeek)
        & \SENLGPearsonDeepSeekMedium (\SENLGPearsonPValueDeepSeekMedium)
        & \SENLGPearsonDeepSeekHard (\SENLGPearsonPValueDeepSeekHard) \\

      \SEEmbed
      & \SEEmbedPearsonGPT (\SEEmbedPearsonPValueGPT)
      & \SEEmbedPearsonGPTMedium (\SEEmbedPearsonPValueGPTMedium)
      & \SEEmbedPearsonGPTHard (\SEEmbedPearsonPValueGPTHard)
      & \SEEmbedPearsonDeepSeek (\SEEmbedPearsonPValueDeepSeek)
      & \SEEmbedPearsonDeepSeekMedium (\SEEmbedPearsonPValueDeepSeekMedium)
      & \SEEmbedPearsonDeepSeekHard (\SEEmbedPearsonPValueDeepSeekHard) \\

      \SECodeBLEU
        & \SENLGCBPearsonGPT (\SENLGCBPearsonPValueGPT)
        & \SENLGCBPearsonGPTMedium (\SENLGCBPearsonPValueGPTMedium)
        & \SENLGCBPearsonGPTHard (\SENLGCBPearsonPValueGPTHard)
        & \SENLGCBPearsonDeepSeek (\SENLGCBPearsonPValueDeepSeek)
        & \SENLGCBPearsonDeepSeekMedium (\SENLGCBPearsonPValueDeepSeekMedium)
        & \SENLGCBPearsonDeepSeekHard (\SENLGCBPearsonPValueDeepSeekHard) \\

      \SESymbolic
        & \textbf{\SESymbolicPearsonGPTUpdated (\SESymbolicPearsonPValueGPTUpdated)}
        & \textbf{\SESymbolicPearsonGPTMediumUpdated (\SESymbolicPearsonPValueGPTMediumUpdated)}
        & \textbf{\SESymbolicPearsonGPTHardUpdated (\SESymbolicPearsonPValueGPTHardUpdated)}
        & \textbf{\SESymbolicPearsonDeepSeek (\SESymbolicPearsonPValueDeepSeek)}
        & \textbf{\SESymbolicPearsonDeepSeekMedium (\SESymbolicPearsonPValueDeepSeekMedium)}
        & \textbf{\SESymbolicPearsonDeepSeekHard (\SESymbolicPearsonPValueDeepSeekHard)} \\

      \SESymbolicUnif
        & \textbf{\SESymbolicUnifPearsonGPTUpdated (\SESymbolicUnifPearsonPValueGPTUpdated)}
        & \textbf{\SESymbolicUnifPearsonGPTMediumUpdated (\SESymbolicUnifPearsonPValueGPTMediumUpdated)}
        & \textbf{\SESymbolicUnifPearsonGPTHardUpdated (\SESymbolicUnifPearsonPValueGPTHardUpdated)}
        & \textbf{\SESymbolicUnifPearsonDeepSeek (\SESymbolicUnifPearsonPValueDeepSeek)}
        & \textbf{\SESymbolicUnifPearsonDeepSeekMedium (\SESymbolicUnifPearsonPValueDeepSeekMedium)}
        & \textbf{\SESymbolicUnifPearsonDeepSeekHard (\SESymbolicUnifPearsonPValueDeepSeekHard)} \\

      \midrule

      \MIOriginal
        & \MINLGPearsonGPTUpdated (\MINLGPearsonPValueGPTUpdated)
        & \MINLGPearsonGPTMediumUpdated (\MINLGPearsonPValueGPTMediumUpdated)
        & \MINLGPearsonGPTHardUpdated (\MINLGPearsonPValueGPTHardUpdated)
        & \MINLGPearsonDeepSeek (\MINLGPearsonPValueDeepSeek)
        & \MINLGPearsonDeepSeekMedium (\MINLGPearsonPValueDeepSeekMedium)
        & \MINLGPearsonDeepSeekHard (\MINLGPearsonPValueDeepSeekHard) \\
      \MIEmbed
      & \MIEmbedPearsonGPT (\MIEmbedPearsonPValueGPT)
      & \MIEmbedPearsonGPTMedium (\MIEmbedPearsonPValueGPTMedium)
      & \MIEmbedPearsonGPTHard (\MIEmbedPearsonPValueGPTHard)
      & \MIEmbedPearsonDeepSeek (\MIEmbedPearsonPValueDeepSeek)
      & \MIEmbedPearsonDeepSeekMedium (\MIEmbedPearsonPValueDeepSeekMedium)
      & \MIEmbedPearsonDeepSeekHard (\MIEmbedPearsonPValueDeepSeekHard) \\

      \MICodeBLEU
        & \MINLGCBPearsonGPT (\MINLGCBPearsonPValueGPT)
        & \MINLGCBPearsonGPTMedium (\MINLGCBPearsonPValueGPTMedium)
        & \MINLGCBPearsonGPTHard (\MINLGCBPearsonPValueGPTHard)
        & \MINLGCBPearsonDeepSeek (\MINLGCBPearsonPValueDeepSeek)
        & \MINLGCBPearsonDeepSeekMedium (\MINLGCBPearsonPValueDeepSeekMedium)
        & \MINLGCBPearsonDeepSeekHard (\MINLGCBPearsonPValueDeepSeekHard) \\

      \MISymbolic
        & \textbf{\MISymbolicPearsonGPTUpdated (\MISymbolicPearsonPValueGPTUpdated)}
        & \textbf{\MISymbolicPearsonGPTMediumUpdated (\MISymbolicPearsonPValueGPTMediumUpdated)}
        & \textbf{\MISymbolicPearsonGPTHardUpdated (\MISymbolicPearsonPValueGPTHardUpdated)}
        & \textbf{\MISymbolicPearsonDeepSeek (\MISymbolicPearsonPValueDeepSeek)}
        & \textbf{\MISymbolicPearsonDeepSeekMedium (\MISymbolicPearsonPValueDeepSeekMedium)}
        & \textbf{\MISymbolicPearsonDeepSeekHard (\MISymbolicPearsonPValueDeepSeekHard)} \\

      \MISymbolicUnif
      & \textbf{\MISymbolicUnifPearsonGPT (\MISymbolicUnifPearsonPValueGPT)}
      & \textbf{\MISymbolicUnifPearsonGPTMedium (\MISymbolicUnifPearsonPValueGPTMedium)}
      & \textbf{\MISymbolicUnifPearsonGPTHard (\MISymbolicUnifPearsonPValueGPTHard)}
      & \textbf{\MISymbolicUnifPearsonDeepSeek (\MISymbolicUnifPearsonPValueDeepSeek)}
      & \textbf{\MISymbolicUnifPearsonDeepSeekMedium (\MISymbolicUnifPearsonPValueDeepSeekMedium)}
      & \textbf{\MISymbolicUnifPearsonDeepSeekHard (\MISymbolicUnifPearsonPValueDeepSeekHard)} \\

      \midrule

        \CCEmbedding
        & \CCEmbedPearsonGPT (\CCEmbedPearsonPValueGPT)
        & \CCEmbedPearsonGPTMedium (\CCEmbedPearsonPValueGPTMedium)
        & \CCEmbedPearsonGPTHard (\CCEmbedPearsonPValueGPTHard)
        & \CCEmbedPearsonDeepSeek (\CCEmbedPearsonPValueDeepSeek)
        & \CCEmbedPearsonDeepSeekMedium (\CCEmbedPearsonPValueDeepSeekMedium)
        & \CCEmbedPearsonDeepSeekHard (\CCEmbedPearsonPValueDeepSeekHard) \\

      \CCSymbolic
        & \textbf{\CCSymbolicPearsonGPT (\CCSymbolicPearsonPValueGPT)}
        & \textbf{\CCSymbolicPearsonGPTMedium (\CCSymbolicPearsonPValueGPTMedium)}
        & \textbf{\CCSymbolicPearsonGPTHard (\CCSymbolicPearsonPValueGPTHard)}
        & \textbf{\CCSymbolicPearsonDeepSeek (\CCSymbolicPearsonPValueDeepSeek)}
        & \textbf{\CCSymbolicPearsonDeepSeekMedium (\CCSymbolicPearsonPValueDeepSeekMedium)}
        & \textbf{\CCSymbolicPearsonDeepSeekHard (\CCSymbolicPearsonPValueDeepSeekHard)} \\
      \midrule

      \LLMProbability
        & \LLMProbabilityPearsonGPTUpdated (\LLMProbabilityPearsonPValueGPTUpdated)
        & \LLMProbabilityPearsonGPTMediumUpdated (\LLMProbabilityPearsonPValueGPTMediumUpdated)
        & \LLMProbabilityPearsonGPTHardUpdated (\LLMProbabilityPearsonPValueGPTHardUpdated)
        & \LLMProbabilityPearsonDeepSeek (\LLMProbabilityPearsonPValueDeepSeek)
        & \LLMProbabilityPearsonDeepSeekMedium(\LLMProbabilityPearsonPValueDeepSeekMedium)
        & \LLMProbabilityPearsonDeepSeekHard (\LLMProbabilityPearsonPValueDeepSeekHard) \\

      \bottomrule
  \end{tabular}
\end{table*}

\vspace{-.2cm}
\subsection{Discussion of Results}\label{sec:results-discussion}
\vspace{-.05cm}


\noindent\textbf{Symbolic clustering enables NLG-based uncertainty estimation methods to perform effectively in code generation.}
As shown in Table~\ref{tab:correlation_results}, the \SE and \MI techniques fail to exhibit statistically significant correlation between estimated uncertainty and correctness when applied to code generation---even when adapted with OpenAI's \gpt\gptembed embeddings, which are tailored for code~\cite{openai2022embedding} or heuristic similarity metrics such as \codebleu{}. Notably, both techniques only achieve statistically significant correlation when paired with symbolic clustering (\SESymbolic, \SESymbolicUnif, \MISymbolic, and \MISymbolicUnif), highlighting the importance of precise semantic equivalence analysis in this domain.

\noindent{\textbf{Semantic equivalence among the LLM responses is a strong indicator of correctness.}}
When relying on log-probabilities computed by the LLM (row corresponding to \LLMProbability in Table~\ref{tab:correlation_results}), we find no statistically significant correlation with correctness, indicating that token-level probabilities are insufficient proxies for correctness in code generation. In contrast, symbolic clustering provides a much stronger signal. Even when assuming a uniform distribution over LLM responses, \SESymbolicUnif and \MISymbolicUnif still yield a statistically significant correlation with correctness.

\noindent\textbf{Lightweight metrics based on symbolic clustering can still yield effective correctness estimates.}
Pushing this abstraction further, \CCSymbolic, which relies solely on the number of symbolic clusters, also correlates with correctness.
Crucially, this correlation hinges on the precision of clustering: \CC-\codeembed doesn't exhibit statistically significant results.
This result is particularly relevant because many modern LLMs do not expose reliable log-probabilities. For instance, it allows us to evaluate \CCSymbolic and \SESymbolicUnif---methods that disregard the LLM log-probabilities--- for \claude (results in Table~\ref{tab:claude_symbolic_correlation}, Appendix), showing that both achieve a statistically significant correlation with correctness.

Our results hold across different models and different problem complexities.

\noindent\textbf{Abstention policies.} 
We envision uncertainty scores being utilised as part of an abstention policy, similar to the approach described by Abbasi~\etal~\cite{abbasi2024believe}.
In this policy, an uncertainty threshold is set such that LLM responses with scores below the threshold are assumed to be functionally corrent, while those above the threshold are considered incorrect. For the latter, the LLM may abstain from presenting these responses to the user. LLM responses are classified as correct if their correctness score exceeds a predefined threshold and incorrect otherwise. We used a correctness score threshold of 90\%. To determine the corresponding \emph{uncertainty threshold} for the abstention policy, we select the value that maximises correctness accuracy.
We split the dataset 50/50 into training and validation.
To mitigate overfitting and reduce bias,  we employed 2-fold cross-validation, alternating between training and validation for each fold.

%

For our experimental evaluation, we focus on the best-performing uncertainty estimation techniques—\SESymbolic, \SESymbolicUnif, and \CCSymbolic—while using \SEEmbed and \LLMProbability as baselines for comparison.
Table~\ref{tab:abstention_metrics} shows that \SESymbolic, \SESymbolicUnif, and \CCSymbolic, all using symbolic clustering, achieve high accuracy and extremely low false positive rates, i.e. the fraction of incorrect responses mistakenly shown to the user. This conservative acceptance of LLM outputs is crucial for code generation, where incorrect code poses safety risks.
False negative rates (i.e., correct responses unnecessarily filtered out) range from 11.9\% to 20.3\%, an acceptable trade-off for enhanced reliability.


\SEEmbed and \LLMProbability performed significantly worse across accuracy, false positive, and false negative rates, rendering them ineffective in practice. In particular, \SEEmbed’s poor results highlight that simply replacing NLG-specific embeddings with code-specific ones is insufficient for adapting uncertainty estimation techniques to code generation and building effective abstention policies.

\begin{table*}[ht!]
  \centering
  \caption{Abstention metrics} 
  \label{tab:abstention_metrics}
  \footnotesize
  \begin{tabular}{@{}l  r r r   r r r @{}}
    \toprule
    \multirow{2}{*}{\textbf{Technique}}
      & \multicolumn{3}{c}{\textbf{\gptturbo}}
      & \multicolumn{3}{c}{\textbf{\deepseek}} \\
    \cmidrule(lr){2-4}\cmidrule(lr){5-7}
      & \textbf{Accuracy}
      & \textbf{False Pos.}
      & \textbf{False Neg.}
      & \textbf{Accuracy}
      & \textbf{False Pos.}
      & \textbf{False Neg.} \\
    \midrule
    \SESymbolic
      & \SENormAcc
      & \SENormFP
      & \SENormFN
      & \SENormAccDS      
      & \SENormFPDS       
      & \SENormFNDS       
      \\
    \SESymbolicUnif
      & \SEUnifAcc
      & \SEUnifFP
      & \SEUnifFN
      & \SEUnifAccDS      
      & \SEUnifFPDS       
      & \SEUnifFNDS       
      \\
      \CCSymbolic
      & \CCSymbAcc
      & \CCSymbFP
      & \CCSymbFN
      & \CCSymbAccDS  
      & \CCSymbFPDS   
      & \CCSymbFNDS   
      \\
      \SEEmbed
      & \SEEmbedAcc
      & \SEEmbedFP
      & \SEEmbedFN
      & \SEEmbedAccDS      
      & \SEEmbedFPDS       
      & \SEEmbedFNDS       
      \\
    \LLMProbability
      & \LLMProbabilityAcc
      & \LLMProbabilityFP
      & \LLMProbabilityFN
      & \LLMProbabilityAccDS  
      & \LLMProbabilityFPDS   
      & \LLMProbabilityFNDS   
      \\
    \bottomrule
  \end{tabular}
\end{table*}

\noindent\textbf{Limitations.} A key limitation of our approach lies in the bounded nature of symbolic execution, particularly with respect to loops. To ensure tractability, we unroll loops and explore program paths only up to the timeout described in the experimental setup. While this enables efficient analysis, it introduces the risk of false positives in equivalence checking: two code snippets may be deemed semantically equivalent simply because their behavioural differences only manifest beyond the unrolling limit.
Another important limitation is that we only focus on the NLG-based uncertainty estimation techniques proposed by Kuhn~\etal~\cite{kuhnsemantic,farquhar2024detecting} and Abbasi~\etal~\cite{abbasi2024believe}. We mitigated this by selecting two recent state-of-the-art methods that have received significant attention and citations in the literature.

\section{Related Work}
\label{sec:related}
Numerous approaches have been proposed for LLM evaluation in NLG. One class of methods involves training dedicated evaluators~\cite{DBLP:journals/corr/abs-2306-05087,DBLP:journals/corr/abs-2308-04592} or using proprietary LLMs such as ChatGPT~\cite{DBLP:journals/corr/abs-2303-04048,DBLP:conf/eamt/KocmiF23}. However, these approaches are static, opaque, or biased~\cite{DBLP:conf/acl/WangLCCZLCKLLS24}, and are outside the scope of our oracle-free setting.
Within NLG, oracle-free evaluation has leveraged both self-assessment~\cite{DBLP:conf/icml/0001TDM24,DBLP:conf/acl/Xia0WCZ24,DBLP:journals/tmlr/LinHE22,DBLP:journals/corr/abs-2207-05221} and internal signals—such as token-level log-probabilities—to estimate uncertainty~\cite{abbasi2024believe,farquhar2024detecting,kuhnsemantic,DBLP:conf/iclr/MalininG21}.

In this paper, we focus on the most recent state-of-the-art approaches based on entropy~\cite{farquhar2024detecting,kuhnsemantic} and mutual information~\cite{abbasi2024believe}, which have demonstrated strong performance in NLG.
Other techniques~\cite{lahlou2021deup,desai2020calibration,mielke2022reducing,osband2023epistemic,cole2023selectively,yona2024narrowing,hou2023decomposing,DBLP:journals/corr/abs-2207-05221} are either shown to be less effective or rely on costly additional training steps. In another related work, Jiang et al.~\cite{jiang2021can} show that naive log-likelihoods are weak correctness proxies.


While all the aforementioned works focus on natural language, most code generation studies rely on external oracles to assess the quality of LLM-generated code.
For example, works such as \cite{liu2024your} and \cite{vaithilingam2022expectation} conduct large-scale evaluations of LLM-generated code against existing correctness oracles. Similarly, Honarvar~\etal~\cite{honarvar2024turbulencesystematicallyautomaticallytesting} introduce a test suite expansion technique to generate additional code generation problems with oracles, enabling a more comprehensive evaluation of LLM performance.

Another closely related area is hallucination detection, where studies such as \cite{liu2024exploring} and \cite{eghbali2024hallucinator} focus on evaluating LLMs using problems specifically designed to expose inconsistencies and hallucinations.
Liu et al.\cite{liu2024exploring} establish a taxonomy of hallucinations in LLM-generated code and propose a benchmark to evaluate the ability of code LLMs to recognise hallucinations. 
Eghbali and Pradel\cite{eghbali2024hallucinator} introduce a technique that grounds LLM predictions by retrieving relevant API references and iteratively refining prompts with increasingly relevant context, leading to more reliable LLM-generated code.

Closest to our work, Huang~\etal~\cite{huang2025look} explore lightweight uncertainty proxies (\eg \textsc{CodeBLEU}) for code generation, but find them ineffective. Unlike prior work, we introduce symbolic clustering and propose metrics that show correlation with correctness---making us, to our knowledge, the first to identify effective oracle-free uncertainty estimators in code generation.

\section{Conclusion}
\label{sec:conclusion}
Our investigation reveals that existing uncertainty estimation techniques from natural language generation fail to generalise to code generation unless paired with precise semantic clustering. By introducing symbolic execution–based clustering, we restore the correlation between estimated uncertainty and correctness. This enables both more accurate uncertainty estimation and the definition of abstention policies with near-zero false positives. We further demonstrate that a simple symbolic cluster count offers a lightweight yet effective proxy for correctness, providing practical value even in settings where token-level probabilities are unavailable.


\begingroup
  \small               
  \bibliographystyle{plainnat}  
  \bibliography{ref}           

\begin{thebibliography}{53}
\providecommand{\natexlab}[1]{#1}
\providecommand{\url}[1]{\texttt{#1}}
\expandafter\ifx\csname urlstyle\endcsname\relax
  \providecommand{\doi}[1]{doi: #1}\else
  \providecommand{\doi}{doi: \begingroup \urlstyle{rm}\Url}\fi

\bibitem[ope()]{openai}
{OpenAI}.
\newblock \url{https://openai.com}.
\newblock Accessed: January 19, 2025.

\bibitem[gpt(2023)]{gpt35turboinstruct}
{gpt-3.5-turbo-instruct}.
\newblock \url{https://platform.openai.com/docs/models/gpt-3-5}, 2023.
\newblock Accessed: January 19, 2025.

\bibitem[Abbasi-Yadkori et~al.(2024)Abbasi-Yadkori, Kuzborskij, Gy{\"o}rgy, and
  Szepesvari]{abbasi2024believe}
Yasin Abbasi-Yadkori, Ilja Kuzborskij, Andr{\'a}s Gy{\"o}rgy, and Csaba
  Szepesvari.
\newblock To believe or not to believe your llm: Iterative prompting for
  estimating epistemic uncertainty.
\newblock In \emph{The Thirty-eighth Annual Conference on Neural Information
  Processing Systems}, 2024.

\bibitem[{Anthropic}(2023)]{anthropic2023claude3}
{Anthropic}.
\newblock Claude 3: A large language model.
\newblock \url{https://www.anthropic.com/claude}, 2023.

\bibitem[Bruni et~al.(2011)Bruni, Disney, and Flanagan]{Bruni2011APA}
Alessandro~Maria Bruni, Tim Disney, and Cormac Flanagan.
\newblock A peer architecture for lightweight symbolic execution.
\newblock 2011.
\newblock URL \url{https://api.semanticscholar.org/CorpusID:61565931}.

\bibitem[Cadar et~al.(2008)Cadar, Dunbar, and Engler]{symex_klee}
Cristian Cadar, Daniel Dunbar, and Dawson Engler.
\newblock Klee: unassisted and automatic generation of high-coverage tests for
  complex systems programs.
\newblock In \emph{Proceedings of the 8th USENIX Conference on Operating
  Systems Design and Implementation}, OSDI'08, page 209–224, USA, 2008.
  USENIX Association.

\bibitem[Chaney(2023)]{crosshair}
Phillip~S. Chaney.
\newblock Crosshair: A python analysis tool for exploring the correctness of
  programs.
\newblock \url{https://github.com/pschanely/CrossHair}, 2023.
\newblock Accessed: January 19, 2025.

\bibitem[Cole et~al.(2023)Cole, Zhang, Gillick, Eisenschlos, Dhingra, and
  Eisenstein]{cole2023selectively}
Jeremy~R Cole, Michael~JQ Zhang, Daniel Gillick, Julian~Martin Eisenschlos,
  Bhuwan Dhingra, and Jacob Eisenstein.
\newblock Selectively answering ambiguous questions.
\newblock \emph{arXiv preprint arXiv:2305.14613}, 2023.

\bibitem[David et~al.(2025)David, Kesseli, Kroening, and
  Zhang]{david2025codehints}
Cristina David, Pascal Kesseli, Daniel Kroening, and Hanliang Zhang.
\newblock Quantifying the benefits of code hints for refactoring deprecated
  java apis.
\newblock In \emph{Companion Proceedings of the 33rd ACM SIGSOFT International
  Symposium on the Foundations of Software Engineering (FSE 2025 - Industry
  Track)}, Trondheim, Norway, 2025. ACM.
\newblock Industry Track.

\bibitem[Desai and Durrett(2020)]{desai2020calibration}
Shrey Desai and Greg Durrett.
\newblock Calibration of pre-trained transformers.
\newblock \emph{arXiv preprint arXiv:2003.07892}, 2020.

\bibitem[Eghbali and Pradel(2024)]{eghbali2024hallucinator}
Aryaz Eghbali and Michael Pradel.
\newblock De-hallucinator: Iterative grounding for llm-based code completion.
\newblock \emph{arXiv preprint arXiv:2401.01701}, 2024.

\bibitem[Eniser et~al.(2024)Eniser, Zhang, David, Wang, Christakis, Paulsen,
  Dodds, and Kroening]{codetranslation2}
Hasan~Ferit Eniser, Hanliang Zhang, Cristina David, Meng Wang, Maria
  Christakis, Brandon Paulsen, Joey Dodds, and Daniel Kroening.
\newblock Towards translating real-world code with llms: A study of translating
  to rust.
\newblock \emph{arXiv preprint arXiv:2405.11514}, 2024.

\bibitem[Fan et~al.(2023)Fan, Gao, Mirchev, Roychoudhury, and
  Tan]{AutoCodeRoverPre}
Zhiyu Fan, Xiang Gao, Martin Mirchev, Abhik Roychoudhury, and Shin~Hwei Tan.
\newblock Automated repair of programs from large language models.
\newblock In \emph{Proceedings of the 45th International Conference on Software
  Engineering}, ICSE '23, page 1469–1481. IEEE Press, 2023.
\newblock ISBN 9781665457019.
\newblock \doi{10.1109/ICSE48619.2023.00128}.
\newblock URL \url{https://doi.org/10.1109/ICSE48619.2023.00128}.

\bibitem[Farquhar et~al.(2024)Farquhar, Kossen, Kuhn, and
  Gal]{farquhar2024detecting}
Sebastian Farquhar, Jannik Kossen, Lorenz Kuhn, and Yarin Gal.
\newblock Detecting hallucinations in large language models using semantic
  entropy.
\newblock \emph{Nature}, 630\penalty0 (8017):\penalty0 625--630, 2024.

\bibitem[Guo et~al.(2025)Guo, Yang, Zhang, Song, Zhang, Xu, Zhu, Ma, Wang, Bi,
  et~al.]{deepseek}
Daya Guo, Dejian Yang, Haowei Zhang, Junxiao Song, Ruoyu Zhang, Runxin Xu,
  Qihao Zhu, Shirong Ma, Peiyi Wang, Xiao Bi, et~al.
\newblock Deepseek-r1: Incentivizing reasoning capability in llms via
  reinforcement learning.
\newblock \emph{arXiv preprint arXiv:2501.12948}, 2025.

\bibitem[He et~al.(2020)He, Liu, Gao, and Chen]{he2020deberta}
Pengcheng He, Xiaodong Liu, Jianfeng Gao, and Weizhu Chen.
\newblock Deberta: Decoding-enhanced bert with disentangled attention.
\newblock \emph{arXiv preprint arXiv:2006.03654}, 2020.

\bibitem[Honarvar et~al.(2024)Honarvar, van~der Wilk, and
  Donaldson]{honarvar2024turbulencesystematicallyautomaticallytesting}
Shahin Honarvar, Mark van~der Wilk, and Alastair Donaldson.
\newblock Turbulence: Systematically and automatically testing
  instruction-tuned large language models for code, 2024.
\newblock URL \url{https://arxiv.org/abs/2312.14856}.

\bibitem[Hou et~al.(2023)Hou, Liu, Qian, Andreas, Chang, and
  Zhang]{hou2023decomposing}
Bairu Hou, Yujian Liu, Kaizhi Qian, Jacob Andreas, Shiyu Chang, and Yang Zhang.
\newblock Decomposing uncertainty for large language models through input
  clarification ensembling.
\newblock \emph{arXiv preprint arXiv:2311.08718}, 2023.

\bibitem[Huang et~al.(2025)Huang, Song, Wang, Zhao, Chen, Juefei-Xu, and
  Ma]{huang2025look}
Yuheng Huang, Jiayang Song, Zhijie Wang, Shengming Zhao, Huaming Chen, Felix
  Juefei-Xu, and Lei Ma.
\newblock Look before you leap: An exploratory study of uncertainty analysis
  for large language models.
\newblock \emph{IEEE Transactions on Software Engineering}, 2025.

\bibitem[Jain et~al.(2024)Jain, Han, Gu, Li, Yan, Zhang, Wang, Solar-Lezama,
  Sen, and Stoica]{livecodebench}
Naman Jain, King Han, Alex Gu, Wen-Ding Li, Fanjia Yan, Tianjun Zhang, Sida
  Wang, Armando Solar-Lezama, Koushik Sen, and Ion Stoica.
\newblock Livecodebench: Holistic and contamination free evaluation of large
  language models for code, 2024.
\newblock URL \url{https://arxiv.org/abs/2403.07974}.

\bibitem[Jiang et~al.(2021)Jiang, Araki, Ding, and Neubig]{jiang2021can}
Zhengbao Jiang, Jun Araki, Haibo Ding, and Graham Neubig.
\newblock How can we know when language models know? on the calibration of
  language models for question answering.
\newblock \emph{Transactions of the Association for Computational Linguistics},
  9:\penalty0 962--977, 2021.

\bibitem[Johnson et~al.(2024)Johnson, Tarlow, Duvenaud, and
  Maddison]{DBLP:conf/icml/0001TDM24}
Daniel~D. Johnson, Daniel Tarlow, David Duvenaud, and Chris~J. Maddison.
\newblock Experts don't cheat: Learning what you don't know by predicting
  pairs.
\newblock In \emph{Forty-first International Conference on Machine Learning,
  {ICML} 2024, Vienna, Austria, July 21-27, 2024}. OpenReview.net, 2024.
\newblock URL \url{https://openreview.net/forum?id=AVEc9LvSlO}.

\bibitem[Kadavath et~al.(2022)Kadavath, Conerly, Askell, Henighan, Drain,
  Perez, Schiefer, Hatfield{-}Dodds, DasSarma, Tran{-}Johnson, Johnston, Showk,
  Jones, Elhage, Hume, Chen, Bai, Bowman, Fort, Ganguli, Hernandez, Jacobson,
  Kernion, Kravec, Lovitt, Ndousse, Olsson, Ringer, Amodei, Brown, Clark,
  Joseph, Mann, McCandlish, Olah, and
  Kaplan]{DBLP:journals/corr/abs-2207-05221}
Saurav Kadavath, Tom Conerly, Amanda Askell, Tom Henighan, Dawn Drain, Ethan
  Perez, Nicholas Schiefer, Zac Hatfield{-}Dodds, Nova DasSarma, Eli
  Tran{-}Johnson, Scott Johnston, Sheer~El Showk, Andy Jones, Nelson Elhage,
  Tristan Hume, Anna Chen, Yuntao Bai, Sam Bowman, Stanislav Fort, Deep
  Ganguli, Danny Hernandez, Josh Jacobson, Jackson Kernion, Shauna Kravec,
  Liane Lovitt, Kamal Ndousse, Catherine Olsson, Sam Ringer, Dario Amodei, Tom
  Brown, Jack Clark, Nicholas Joseph, Ben Mann, Sam McCandlish, Chris Olah, and
  Jared Kaplan.
\newblock Language models (mostly) know what they know.
\newblock \emph{CoRR}, abs/2207.05221, 2022.
\newblock \doi{10.48550/ARXIV.2207.05221}.
\newblock URL \url{https://doi.org/10.48550/arXiv.2207.05221}.

\bibitem[King(1976)]{king76_symex}
James~C. King.
\newblock Symbolic execution and program testing.
\newblock \emph{Commun. ACM}, 19\penalty0 (7):\penalty0 385–394, July 1976.
\newblock ISSN 0001-0782.
\newblock \doi{10.1145/360248.360252}.
\newblock URL \url{https://doi.org/10.1145/360248.360252}.

\bibitem[Kocmi and Federmann(2023)]{DBLP:conf/eamt/KocmiF23}
Tom Kocmi and Christian Federmann.
\newblock Large language models are state-of-the-art evaluators of translation
  quality.
\newblock In Mary Nurminen, Judith Brenner, Maarit Koponen, Sirkku Latomaa,
  Mikhail Mikhailov, Frederike Schierl, Tharindu Ranasinghe, Eva Vanmassenhove,
  Sergi~Alvarez Vidal, Nora Aranberri, Mara Nunziatini, Carla~Parra
  Escart{\'{\i}}n, Mikel~L. Forcada, Maja Popovic, Carolina Scarton, and Helena
  Moniz, editors, \emph{Proceedings of the 24th Annual Conference of the
  European Association for Machine Translation, {EAMT} 2023, Tampere, Finland,
  12-15 June 2023}, pages 193--203. European Association for Machine
  Translation, 2023.
\newblock URL \url{https://aclanthology.org/2023.eamt-1.19}.

\bibitem[Koehn and Knowles(2017)]{DBLP:conf/aclnmt/KoehnK17}
Philipp Koehn and Rebecca Knowles.
\newblock Six challenges for neural machine translation.
\newblock In Thang Luong, Alexandra Birch, Graham Neubig, and Andrew~M. Finch,
  editors, \emph{Proceedings of the First Workshop on Neural Machine
  Translation, NMT@ACL 2017, Vancouver, Canada, August 4, 2017}, pages 28--39.
  Association for Computational Linguistics, 2017.
\newblock \doi{10.18653/V1/W17-3204}.
\newblock URL \url{https://doi.org/10.18653/v1/w17-3204}.

\bibitem[Kuhn et~al.(2023)Kuhn, Gal, and Farquhar]{kuhnsemantic}
Lorenz Kuhn, Yarin Gal, and Sebastian Farquhar.
\newblock Semantic uncertainty: Linguistic invariances for uncertainty
  estimation in natural language generation.
\newblock In \emph{The Eleventh International Conference on Learning
  Representations, {ICLR} 2023, Kigali, Rwanda, May 1-5, 2023}. OpenReview.net,
  2023.
\newblock URL \url{https://openreview.net/forum?id=VD-AYtP0dve}.

\bibitem[Lahlou et~al.(2021)Lahlou, Jain, Nekoei, Butoi, Bertin, Rector-Brooks,
  Korablyov, and Bengio]{lahlou2021deup}
Salem Lahlou, Moksh Jain, Hadi Nekoei, Victor~Ion Butoi, Paul Bertin, Jarrid
  Rector-Brooks, Maksym Korablyov, and Yoshua Bengio.
\newblock Deup: Direct epistemic uncertainty prediction.
\newblock \emph{arXiv preprint arXiv:2102.08501}, 2021.

\bibitem[Li et~al.(2024)Li, Parsert, and Polgreen]{synthesis1}
Yixuan Li, Julian Parsert, and Elizabeth Polgreen.
\newblock Guiding enumerative program synthesis with large language models.
\newblock In \emph{International Conference on Computer Aided Verification},
  pages 280--301. Springer, 2024.

\bibitem[Lin et~al.(2022)Lin, Hilton, and Evans]{DBLP:journals/tmlr/LinHE22}
Stephanie Lin, Jacob Hilton, and Owain Evans.
\newblock Teaching models to express their uncertainty in words.
\newblock \emph{Trans. Mach. Learn. Res.}, 2022, 2022.
\newblock URL \url{https://openreview.net/forum?id=8s8K2UZGTZ}.

\bibitem[Liu et~al.(2024{\natexlab{a}})Liu, Liu, Shi, Huang, Wang, Yang, Zhang,
  Li, and Ma]{liu2024exploring}
Fang Liu, Yang Liu, Lin Shi, Houkun Huang, Ruifeng Wang, Zhen Yang, Li~Zhang,
  Zhongqi Li, and Yuchi Ma.
\newblock Exploring and evaluating hallucinations in llm-powered code
  generation.
\newblock \emph{arXiv preprint arXiv:2404.00971}, 2024{\natexlab{a}}.

\bibitem[Liu et~al.(2024{\natexlab{b}})Liu, Xia, Wang, and Zhang]{liu2024your}
Jiawei Liu, Chunqiu~Steven Xia, Yuyao Wang, and Lingming Zhang.
\newblock Is your code generated by chatgpt really correct? rigorous evaluation
  of large language models for code generation.
\newblock \emph{Advances in Neural Information Processing Systems}, 36,
  2024{\natexlab{b}}.

\bibitem[Luo et~al.(2024)Luo, Yu, Zhang, Liang, and Xiong]{codetranslation1}
Yang Luo, Richard Yu, Fajun Zhang, Ling Liang, and Yongqiang Xiong.
\newblock Bridging gaps in llm code translation: Reducing errors with call
  graphs and bridged debuggers.
\newblock In \emph{Proceedings of the 39th IEEE/ACM International Conference on
  Automated Software Engineering}, pages 2448--2449, 2024.

\bibitem[Malinin and Gales(2021)]{DBLP:conf/iclr/MalininG21}
Andrey Malinin and Mark J.~F. Gales.
\newblock Uncertainty estimation in autoregressive structured prediction.
\newblock In \emph{9th International Conference on Learning Representations,
  {ICLR} 2021, Virtual Event, Austria, May 3-7, 2021}. OpenReview.net, 2021.
\newblock URL \url{https://openreview.net/forum?id=jN5y-zb5Q7m}.

\bibitem[Mielke et~al.(2022)Mielke, Szlam, Dinan, and
  Boureau]{mielke2022reducing}
Sabrina~J Mielke, Arthur Szlam, Emily Dinan, and Y-Lan Boureau.
\newblock Reducing conversational agents' overconfidence through linguistic
  calibration.
\newblock \emph{Transactions of the Association for Computational Linguistics},
  10:\penalty0 857--872, 2022.

\bibitem[Murray and Chiang(2018)]{DBLP:conf/wmt/MurrayC18}
Kenton Murray and David Chiang.
\newblock Correcting length bias in neural machine translation.
\newblock In Ondrej Bojar, Rajen Chatterjee, Christian Federmann, Mark Fishel,
  Yvette Graham, Barry Haddow, Matthias Huck, Antonio Jimeno{-}Yepes, Philipp
  Koehn, Christof Monz, Matteo Negri, Aur{\'{e}}lie N{\'{e}}v{\'{e}}ol,
  Mariana~L. Neves, Matt Post, Lucia Specia, Marco Turchi, and Karin Verspoor,
  editors, \emph{Proceedings of the Third Conference on Machine Translation:
  Research Papers, {WMT} 2018, Belgium, Brussels, October 31 - November 1,
  2018}, pages 212--223. Association for Computational Linguistics, 2018.
\newblock \doi{10.18653/V1/W18-6322}.
\newblock URL \url{https://doi.org/10.18653/v1/w18-6322}.

\bibitem[OpenAI(2022)]{openai2022embedding}
OpenAI.
\newblock New and improved embedding model.
\newblock \url{https://openai.com/index/new-and-improved-embedding-model/},
  December 2022.
\newblock Accessed: 2025-05-13.

\bibitem[Osband et~al.(2023)Osband, Wen, Asghari, Dwaracherla, Ibrahimi, Lu,
  and Van~Roy]{osband2023epistemic}
Ian Osband, Zheng Wen, Seyed~Mohammad Asghari, Vikranth Dwaracherla, Morteza
  Ibrahimi, Xiuyuan Lu, and Benjamin Van~Roy.
\newblock Epistemic neural networks.
\newblock \emph{Advances in Neural Information Processing Systems},
  36:\penalty0 2795--2823, 2023.

\bibitem[Pan et~al.(2024)Pan, Ibrahimzada, Krishna, Sankar, Wassi, Merler,
  Sobolev, Pavuluri, Sinha, and Jabbarvand]{codetranslation3}
Rangeet Pan, Ali~Reza Ibrahimzada, Rahul Krishna, Divya Sankar, Lambert~Pouguem
  Wassi, Michele Merler, Boris Sobolev, Raju Pavuluri, Saurabh Sinha, and
  Reyhaneh Jabbarvand.
\newblock Lost in translation: A study of bugs introduced by large language
  models while translating code.
\newblock In \emph{Proceedings of the IEEE/ACM 46th International Conference on
  Software Engineering}, pages 1--13, 2024.

\bibitem[Pomian et~al.(2024)Pomian, Bellur, Dilhara, Kurbatova, Bogomolov,
  Bryksin, and Dig]{coderefactoring2}
Dorin Pomian, Abhiram Bellur, Malinda Dilhara, Zarina Kurbatova, Egor
  Bogomolov, Timofey Bryksin, and Danny Dig.
\newblock Next-generation refactoring: Combining llm insights and ide
  capabilities for extract method.
\newblock In \emph{2024 IEEE International Conference on Software Maintenance
  and Evolution (ICSME)}, pages 275--287. IEEE, 2024.

\bibitem[Ren et~al.(2020)Ren, Guo, Lu, Zhou, Liu, Tang, Sundaresan, Zhou,
  Blanco, and Ma]{DBLP:journals/corr/abs-2009-10297}
Shuo Ren, Daya Guo, Shuai Lu, Long Zhou, Shujie Liu, Duyu Tang, Neel
  Sundaresan, Ming Zhou, Ambrosio Blanco, and Shuai Ma.
\newblock Codebleu: a method for automatic evaluation of code synthesis.
\newblock \emph{CoRR}, abs/2009.10297, 2020.
\newblock URL \url{https://arxiv.org/abs/2009.10297}.

\bibitem[Shirafuji et~al.(2023)Shirafuji, Oda, Suzuki, Morishita, and
  Watanobe]{coderefactoring1}
Atsushi Shirafuji, Yusuke Oda, Jun Suzuki, Makoto Morishita, and Yutaka
  Watanobe.
\newblock Refactoring programs using large language models with few-shot
  examples.
\newblock In \emph{2023 30th Asia-Pacific Software Engineering Conference
  (APSEC)}, pages 151--160. IEEE, 2023.

\bibitem[Tao et~al.(2024)Tao, Ventresque, Nallur, and Saber]{synthesis2}
Ning Tao, Anthony Ventresque, Vivek Nallur, and Takfarinas Saber.
\newblock Enhancing program synthesis with large language models using
  many-objective grammar-guided genetic programming.
\newblock \emph{Algorithms}, 17\penalty0 (7):\penalty0 287, 2024.

\bibitem[Vaithilingam et~al.(2022)Vaithilingam, Zhang, and
  Glassman]{vaithilingam2022expectation}
Priyan Vaithilingam, Tianyi Zhang, and Elena~L Glassman.
\newblock Expectation vs. experience: Evaluating the usability of code
  generation tools powered by large language models.
\newblock In \emph{Chi conference on human factors in computing systems
  extended abstracts}, pages 1--7, 2022.

\bibitem[Wang et~al.(2023{\natexlab{a}})Wang, Liang, Meng, Shi, Li, Xu, Qu, and
  Zhou]{DBLP:journals/corr/abs-2303-04048}
Jiaan Wang, Yunlong Liang, Fandong Meng, Haoxiang Shi, Zhixu Li, Jinan Xu,
  Jianfeng Qu, and Jie Zhou.
\newblock Is chatgpt a good {NLG} evaluator? {A} preliminary study.
\newblock \emph{CoRR}, abs/2303.04048, 2023{\natexlab{a}}.
\newblock \doi{10.48550/ARXIV.2303.04048}.
\newblock URL \url{https://doi.org/10.48550/arXiv.2303.04048}.

\bibitem[Wang et~al.(2024)Wang, Li, Chen, Cai, Zhu, Lin, Cao, Kong, Liu, Liu,
  and Sui]{DBLP:conf/acl/WangLCCZLCKLLS24}
Peiyi Wang, Lei Li, Liang Chen, Zefan Cai, Dawei Zhu, Binghuai Lin, Yunbo Cao,
  Lingpeng Kong, Qi~Liu, Tianyu Liu, and Zhifang Sui.
\newblock Large language models are not fair evaluators.
\newblock In Lun{-}Wei Ku, Andre Martins, and Vivek Srikumar, editors,
  \emph{Proceedings of the 62nd Annual Meeting of the Association for
  Computational Linguistics (Volume 1: Long Papers), {ACL} 2024, Bangkok,
  Thailand, August 11-16, 2024}, pages 9440--9450. Association for
  Computational Linguistics, 2024.
\newblock \doi{10.18653/V1/2024.ACL-LONG.511}.
\newblock URL \url{https://doi.org/10.18653/v1/2024.acl-long.511}.

\bibitem[Wang et~al.(2023{\natexlab{b}})Wang, Geng, Lin, Sun, Wen, Liu, Li,
  Bissyand\'{e}, and Mao]{codebleu_usage}
Shangwen Wang, Mingyang Geng, Bo~Lin, Zhensu Sun, Ming Wen, Yepang Liu, Li~Li,
  Tegawend\'{e}~F. Bissyand\'{e}, and Xiaoguang Mao.
\newblock Natural language to code: How far are we?
\newblock In \emph{Proceedings of the 31st ACM Joint European Software
  Engineering Conference and Symposium on the Foundations of Software
  Engineering}, ESEC/FSE 2023, page 375–387, New York, NY, USA,
  2023{\natexlab{b}}. Association for Computing Machinery.
\newblock ISBN 9798400703270.
\newblock \doi{10.1145/3611643.3616323}.
\newblock URL \url{https://doi.org/10.1145/3611643.3616323}.

\bibitem[Wang et~al.(2023{\natexlab{c}})Wang, Yu, Tan, O'Brien, Pasunuru,
  Dwivedi{-}Yu, Golovneva, Zettlemoyer, Fazel{-}Zarandi, and
  Celikyilmaz]{DBLP:journals/corr/abs-2308-04592}
Tianlu Wang, Ping Yu, Xiaoqing~Ellen Tan, Sean O'Brien, Ramakanth Pasunuru,
  Jane Dwivedi{-}Yu, Olga Golovneva, Luke Zettlemoyer, Maryam Fazel{-}Zarandi,
  and Asli Celikyilmaz.
\newblock Shepherd: {A} critic for language model generation.
\newblock \emph{CoRR}, abs/2308.04592, 2023{\natexlab{c}}.
\newblock \doi{10.48550/ARXIV.2308.04592}.
\newblock URL \url{https://doi.org/10.48550/arXiv.2308.04592}.

\bibitem[Wang et~al.(2023{\natexlab{d}})Wang, Yu, Zeng, Yang, Wang, Chen,
  Jiang, Xie, Wang, Xie, Ye, Zhang, and
  Zhang]{DBLP:journals/corr/abs-2306-05087}
Yidong Wang, Zhuohao Yu, Zhengran Zeng, Linyi Yang, Cunxiang Wang, Hao Chen,
  Chaoya Jiang, Rui Xie, Jindong Wang, Xing Xie, Wei Ye, Shikun Zhang, and Yue
  Zhang.
\newblock Pandalm: An automatic evaluation benchmark for {LLM} instruction
  tuning optimization.
\newblock \emph{CoRR}, abs/2306.05087, 2023{\natexlab{d}}.
\newblock \doi{10.48550/ARXIV.2306.05087}.
\newblock URL \url{https://doi.org/10.48550/arXiv.2306.05087}.

\bibitem[Xia et~al.(2024)Xia, Yu, Wu, Chang, and Zhou]{DBLP:conf/acl/Xia0WCZ24}
Tingyu Xia, Bowen Yu, Yuan Wu, Yi~Chang, and Chang Zhou.
\newblock Language models can evaluate themselves via probability discrepancy.
\newblock In Lun{-}Wei Ku, Andre Martins, and Vivek Srikumar, editors,
  \emph{Findings of the Association for Computational Linguistics, {ACL} 2024,
  Bangkok, Thailand and virtual meeting, August 11-16, 2024}, pages 4889--4901.
  Association for Computational Linguistics, 2024.
\newblock \doi{10.18653/V1/2024.FINDINGS-ACL.291}.
\newblock URL \url{https://doi.org/10.18653/v1/2024.findings-acl.291}.

\bibitem[Yona et~al.(2024)Yona, Aharoni, and Geva]{yona2024narrowing}
Gal Yona, Roee Aharoni, and Mor Geva.
\newblock Narrowing the knowledge evaluation gap: Open-domain question
  answering with multi-granularity answers.
\newblock \emph{arXiv preprint arXiv:2401.04695}, 2024.

\bibitem[Zhang et~al.(2025)Zhang, David, Meng, Paulsen, and Kroening]{modular}
Hanliang Zhang, Cristina David, Wang Meng, Brandon Paulsen, and Daniel
  Kroening.
\newblock Scalable, validated code translation of entire projects using large
  language models.
\newblock \emph{Programming Language Design and Implementation (PLDI)}, 2025.

\bibitem[Zhang et~al.(2024)Zhang, Ruan, Fan, and Roychoudhury]{AutoCodeRover}
Yuntong Zhang, Haifeng Ruan, Zhiyu Fan, and Abhik Roychoudhury.
\newblock Autocoderover: Autonomous program improvement.
\newblock In \emph{Proceedings of the 33rd ACM SIGSOFT International Symposium
  on Software Testing and Analysis}, ISSTA 2024, page 1592–1604, New York,
  NY, USA, 2024. Association for Computing Machinery.
\newblock ISBN 9798400706127.
\newblock \doi{10.1145/3650212.3680384}.
\newblock URL \url{https://doi.org/10.1145/3650212.3680384}.

\end{thebibliography}
\endgroup


\appendix


\section{Semantic Uncertainty via Symbolic Clustering}
\label{sec:symex}
This section presents our adaptation of the semantic entropy-based approach by Kuhn~\etal~\cite{kuhnsemantic} for code generation. 
We follow the main steps from the original work while diverging in two key aspects: the way we estimate the distribution of generated LLM responses and the clustering methodology.


\subsubsection{Generation}
The first step involves sampling $M$ code snippets (using the same hyperparameters as Kuhn~\etal~\cite{kuhnsemantic}), $\{s^{(1)}, \ldots, s^{(M)}\}$, from the LLM's output distribution $p(s \mid x)$ for a given prompt $x$. 
The probabilities of the collected samples are processed using a softmax-style normalization function, ensuring that the resulting values can be interpreted as a valid probability distribution.

Given that this process can lead to numerical underflows, to mitigate this, we approximate the probability distribution of the LLM responses using either length-normalization or a uniform distribution.
Following this approximation, let \(\tilde{p}(s \mid x)\) denote the probability of a snippet \(s\) according to the adjusted distribution.




\subsubsection{Clustering via Symbolic Execution}
The second step works by grouping the aforementioned snippets into clusters based on semantic equivalence. 
%
This process, as shown in Algorithm~\ref{alg:clustering} from Section~\ref{sec:symexclustering}, is based on symbolic execution. 

\subsubsection{Entropy Estimation}
The final step computes uncertainty as the semantic entropy over clusters, reflecting the diversity of functional behaviours.

%
First, the probability associated with a cluster \(c\) is calculated as:
\begin{equation*}
    \tilde{p}(c \mid x) = \sum_{s \in c} \tilde{p}(s \mid x),
\end{equation*}
where \(s \in c\) indicates that the snippet \(s\) belongs to the cluster \(c\).
Then, the entropy \(H(C \mid x)\) over the set of clusters \(C\) is defined as:
\begin{equation*}
    H(C \mid x) = -\sum_{c \in C} \log \tilde{p}(c \mid x),
\end{equation*}
where \(C\) denotes all semantic clusters obtained from Algorithm~\ref{alg:clustering} in Section~\ref{sec:symexclustering}. 
%
A higher entropy indicates greater semantic diversity and hence higher uncertainty in the functional behaviour captured by the clusters. 
Conversely, a lower entropy suggests that the model's outputs are concentrated around a few semantically equivalent behaviours, reflecting higher confidence.

\section{Mutual Information Estimation via Symbolic Clustering}
\label{sec:mi}
This section presents an adaptation of the mutual information-based approach for quantifying epistemic uncertainty by Abbasi~\etal~\cite{abbasi2024believe} to the domain of code generation.
We follow the steps from the original work: iterative prompting for generating LLM responses, clustering, and mutual information estimation. 
However, similar to Section~\ref{sec:symex}, we diverge with respect to the methodology for clustering responses and the way we estimate the distribution of generated LLM responses.

\subsubsection{Iterative Prompting for Code Generation}
Iterative prompting is used for generating multiple responses from the LLM and consequently in constructing a pseudo joint distribution of outputs. 


More precisely, the LLM is sampled to produce $n$ responses while also getting their respective probabilities, $\mu(X_j)$ for \(j = 1, 2, \ldots, n\).
These responses are first used to construct iterative prompts by appending the response to the original prompt and asking the LLM to produce more responses. 
This step then makes use of softmax-style normalization to obtain values that can then be treated as probabilities which are used in the subsequent steps.  

As before since this process can lead to numerical underflows, to mitigate this, we use length-normalization.
As opposed to the approach in Section~\ref{sec:symex}, here we did not use the uniform distribution approximation, as the actual LLM-reported probabilities are needed to distinguish between aleatoric and epistemic uncertainties.

Following length normalization, we compute conditional probabilities,  $\mu(X_m|X_n)$ for \(m,n = 1, 2, \ldots, n\), by looking at the response probabilities received from the LLM when subjected to the aforementioned iterative prompts.


\subsubsection{Clustering via Symbolic Execution}
To handle functional diversity, the generated program snippets are clustered based on their semantic equivalence using Algorithm~\ref{alg:clustering}. 


\subsubsection{Mutual Information Estimation}


Once clustering is complete, mutual information is computed over the resulting clusters to quantify epistemic uncertainty. 

The aggregated probabilities are defined as:
\[
\mu_1'(X_i) = \sum_{j \in D(i)} \mu(X_j), \quad
\mu_2'(X_t \mid X_i) = \sum_{j \in D(t)} \mu(X_j \mid X_i),
\]
where \( X_i \) and \( X_t \) are clusters, \( \mu(X_j) \) represents the probability of the output \( X_j \), and \( \mu(X_j \mid X_i) \) is the conditional probability of \( X_j \) given \( X_i \). The set \( D(i) \) contains all outputs assigned to the cluster \( X_i \).

The normalised empirical distributions are:
\[
\hat{\mu}_1(X_i) = \frac{\mu_1'(X_i)}{Z}, \quad \text{where} \quad Z = \sum_{j \in S} \mu_1'(X_j),
\]
\[
\hat{\mu}_2(X_t \mid X_i) = \frac{\mu_2'(X_t \mid X_i)}{Z_i}, \quad \text{where} \quad Z_i = \sum_{j \in S} \mu_2'(X_j \mid X_i).
\]
Here, \( \hat{\mu}_1(X_i) \) is the normalised marginal distribution for cluster \( X_i \), and \( \hat{\mu}_2(X_t \mid X_i) \) is the normalised conditional distribution for \( X_t \) given \( X_i \). The terms \( Z \) and \( Z_i \) are normalization constants to ensure that the distributions sum to 1.

The joint and pseudo-joint distributions are defined as:
\begin{align*}
    \hat{\mu}(X_i, X_t)
      &= \hat{\mu}_1(X_i)\,\hat{\mu}_2(X_t \mid X_i),\\
    \hat{\mu}^\otimes(X_i, X_t)
      &= \hat{\mu}_1(X_i)\,\sum_{j\in S}\hat{\mu}_1(X_j)\,\hat{\mu}_2(X_t \mid X_j).
    \end{align*}
The joint distribution \( \hat{\mu}(X_i, X_t) \) combines the marginal and conditional distributions, while the pseudo-joint distribution \( \hat{\mu}^\otimes(X_i, X_t) \) assumes independence between clusters.

Finally, the mutual information is computed as:
\[
\hat{I}(\gamma_1, \gamma_2) = \sum_{i, t \in S} \hat{\mu}(X_i, X_t) \ln \left( \frac{\hat{\mu}(X_i, X_t) + \gamma_1}{\hat{\mu}^\otimes(X_i, X_t) + \gamma_2} \right).
\]
Here, \( \gamma_1 \) and \( \gamma_2 \) are small stabilization parameters to prevent division by zero, and \( S \) is the set of clusters.



This mutual information score serves as a proxy for epistemic uncertainty. 
High \(\hat{I}\) values signal significant uncertainty.

\section{Additional Evaluation}
\subsection{Experimental Results for \claude}
\begin{table*}[ht!]
  \centering
  \small
  \caption{Correlation Results — Pearson coefficient (p-value) for \claude. Statistically significant results are highlighted in \textbf{bold}.}
  \label{tab:claude_symbolic_correlation}
  \begin{tabular}{@{} l r r r @{}}
    \toprule
    \multirow{2}{*}{\textbf{Technique}}
      & \multicolumn{3}{c}{\textbf{\claude}} \\
    \cmidrule(lr){2-4}
      & \textbf{Easy} & \textbf{Medium} & \textbf{Hard} \\
    \midrule

    \SESymbolicUnif
      & \textbf{\SESymbolicUnifPearsonClaude (\SESymbolicUnifPearsonPValueClaude)}
      & \textbf{\SESymbolicUnifPearsonClaudeMedium (\SESymbolicUnifPearsonPValueClaudeMedium)}
      & \textbf{\SESymbolicUnifPearsonClaudeHard (\SESymbolicUnifPearsonPValueClaudeHard)} \\

    \CCSymbolic
      & \textbf{\CCSymbolicPearsonClaude (\CCSymbolicPearsonPValueClaude)}
      & \textbf{\CCSymbolicPearsonClaudeMedium (\CCSymbolicPearsonPValueClaudeMedium)}
      & \textbf{\CCSymbolicPearsonClaudeHard (\CCSymbolicPearsonPValueClaudeHard)} \\

    \bottomrule
  \end{tabular}
\end{table*}

\subsection{Details on the Abstention Policy Evaluation}
Given that \gptturbo's responses exhibit a low unit test passing rate, leading to an imbalanced dataset where incorrect responses significantly outnumber correct ones. This imbalance risks trivializing the classification task, as an ``all-incorrect'' model would dominate. To address this, we followed standard practice and applied random downsampling, removing a portion of incorrect responses to prevent the model from defaulting to a trivial solution. We split the dataset 50/50 into training and validation.
To mitigate overfitting and reduce bias,  we employed 2-fold cross-validation, alternating between training and validation for each fold.

\end{document}